\documentclass[usenatbib]{mn2e}
\usepackage{epsfig}
\usepackage{amssymb,amsmath,afterpage}

\newcommand{\vv}[1]{\textbf{#1}}	

\newcommand{\Msun}{M_{\odot}}

\begin{document}
\title[Burial of the polar magnetic field of an accreting neutron star]
{Burial of the polar magnetic field of an accreting neutron star. I. Self-consistent analytic and numerical equilibria}

\author[D. J. B. Payne \& A. Melatos]
{D. J. B. Payne$^1$  \& A. Melatos$^1$ \\
$^1$
School of Physics, University of Melbourne,
Parkville, VIC, 3010. Australia. \\
}
\maketitle

\begin{abstract}
The hydromagnetic structure of a neutron star 
accreting symmetrically at both magnetic poles is calculated
as a function of accreted mass, $M_{\rm a}$,
and polar cap radius,
starting from a centered magnetic dipole and evolving
through a quasistatic sequence of two-dimensional,
Grad-Shafranov equilibria.
The calculation is the first to track fully the growth of
high-order magnetic multipoles, due to equatorward hydromagnetic
spreading, while simultaneously preserving
flux freezing and a self-consistent mass-flux distribution.
Equilibria are constructed numerically by an iterative scheme
and analytically by Green functions.
Two key results are obtained,
with implications for recycled pulsars.
(i) The mass required to significantly
reduce the magnetic dipole moment,
$10^{-5}\Msun$,
greatly exceeds previous estimates ($\sim 10^{-10}\Msun$),
which ignored the confining stress
exerted by the
compressed equatorial magnetic field.
(ii) Magnetic bubbles, disconnected from the stellar surface,
form in the later stages of accretion
($M_{\rm a} \gtrsim 10^{-4}\Msun$).
\end{abstract}



\section{INTRODUCTION \label{sec:acc1}}
Observations of low-field neutron stars in binary systems containing
white-dwarf and supergiant companions,
with a history of disc-fed and wind-fed accretion respectively,
suggest that the
magnetic dipole moment $|\vv{m}|$ of a neutron star decreases monotonically
with accreted mass, $M_{\rm a}$
(\citealt{taa86,van95}; see \citealt{wij97} for a dissenting view).
Several mechanisms have been proposed to explain why
$|\vv{m}|$ is reduced:
(i) accelerated Ohmic decay,
where the electrical conductivity of the crust is
lowered by accretion-induced heating
\citep{urp95,urp97};
(ii) interactions between superfluid neutron vortices
and superconducting magnetic
fluxoids in the stellar interior
\citep{mus85,sri90};
and (iii) magnetic screening or burial,
where the currents generating the natal magnetic field are partially
neutralized by accretion-induced currents in the crust
\citep{blo86,aro80}.
For a critical review of these mechanisms,
see \citet{mel01}.

In this paper, we study the mechanism of magnetic burial in detail.
In the early stages of accretion
($M_{\rm a} \lesssim 10^{-10}\Msun$),
accreted matter accumulates 
in a column at the polar cap,
minimally distorting the magnetic field.
The mass-flux distribution in this regime has been
calculated by Grad-Shafranov methods,
with the prediction that $|\vv{m}|$ is reduced by $\sim 1$ per cent
for $M_{\rm a} \approx 10^{-10}\Msun$
\citep{ham83,bro98,lit01}.
We show that these calculations
overestimate the amount of screening; in fact,
$M_{\rm a} \gtrsim 10^{-5}$
is required to reduce $|\vv{m}|$ by 10 per cent
when the confining stress of the compressed,
equatorial magnetic field is modelled faithfully.
In this regime, inaccessible to previous analyses
due to numerical breakdown
\citep{ham83,bro98,lit01},
the latitudinal pressure gradient
at the base of the polar
column forces 
the polar magnetic field to buckle, and the accreted
material spreads equatorward together with
frozen-in magnetic flux \citep{mel01}.
We compute the structure of the
highly distorted magnetic field, and hence $|\vv{m}|$,
as a function of $M_{\rm a}$.

A key advance in the present work is that the
mass-flux distribution in each equilibrium state 
is self-consistent;
our equilibria are generated
by a continuous deformation
of the flux surfaces
of the initial magnetic field (say, a dipole),
in a manner which preserves flux-freezing.
This is not true of previous calculations,
where the mass-flux distribution is unconstrained
relative to the initial state
\citep{bro98,ham83,lit01,mel01}.
However,
several other important effects are not included
to keep the problem manageable.
(i) Ohmic dissipation is neglected, even though the diffusion
and accretion time-scales are comparable for
the smallest magnetic structures predicted
by the theory
\citep{cum01,bro98}.
(ii) We do not investigate the stability of the hydromagnetic
equilibria we compute;
sharp magnetic-field gradients are potentially
disrupted by Rayleigh-Taylor and interchange
instabilities
\citep{cum01,mel01,bha99}.
(iii) We treat the neutron star as a hard surface;
subsidence of accreted material, and incorporation into the
crust, are neglected \citep{bha99}.

The paper is structured as follows.
In Section 2,
we introduce the theoretical framework for calculating
the self-consistent hydromagnetic equilibrium state
of an accreting neutron star.
Analytic and numerical methods of solution are given
in Section 3.
The properties of the equilibria are investigated
in Section 4,
$|\vv{m}|$ is computed as a function of $M_{\rm a}$
and the radius of the polar cap,
and the novel feature of magnetic bubbles is explored.
The limitations of our results, with respect
to time-dependent processes like hydromagnetic
instabilities and ohmic dissipation, are assessed in
Section 5.

\section{THEORY OF EQUILIBRIA}

\subsection{Hydromagnetic force balance}
\label{background}

The equations of non-ideal magnetohydrodynamics (MHD)
in SI units
\citep{ber58}
comprise the equation of mass conservation,
\begin{equation}
\frac{\partial\rho}{\partial t} + \nabla\cdot (\rho\vv{v}) = 0\, ,
\label{continuity}
\end{equation}
the equation of motion,
\begin{equation}
\rho\frac{\partial\vv{v}}{\partial t} + \rho(\vv{v}\cdot\nabla)\vv{v} = -\rho\nabla\phi - \nabla p +\frac{1}{\mu_0}(\nabla\times\vv{B})\times\vv{B}\, ,
\label{mhdforce}
\end{equation}
and the induction equation (minus the displacement current),
\begin{equation}
\frac{\partial\vv{B}}{\partial t} - \nabla\times(\vv{v}\times\vv{B}) = \frac{1}{\mu_0\sigma}\nabla^2\vv{B}\, ,
\label{magevol}
\end{equation}
supplemented by $\nabla\cdot\vv{B} = 0$ and an adiabatic
or isothermal equation of state,
$d(p\rho^{-\Gamma})/dt = 0$.
In this notation, $\vv{B}$,\space$\rho$, \space$p$, \space$\phi$,
\space$\vv{v}$ and $\sigma$
represent the magnetic field, mass density,
kinetic pressure,
gravitational potential,
plasma bulk velocity
and electrical conductivity respectively.
Elastic stresses are neglected
\citep{rom90,mel01},
as is the Hall effect
\citep{gep02}.

In the magnetostatic limit, defined by $\vv{v} = 0$
and $\partial/\partial t = 0$,
the equation of motion reduces to
\begin{equation}
\nabla p + \rho\nabla\phi - \frac{1}{\mu_0}(\nabla\times\vv{B})\times\vv{B} = 0 \space.
\label{mhdforcestatic}
\end{equation}
The local Alfv\'en time-scale,
$\tau_{\rm A} = L/v_{\rm A} \lesssim  4\times 10^{-2} {\rm s}$
for ($L \lesssim 50$ m.),
is much shorter than the accretion time, $\tau_{\rm a}\sim 10^{7}$ yr.
Equations (\ref{continuity}) and (\ref{magevol})
are also satisfied identically (in the ideal-MHD limit
$\sigma\rightarrow\infty$) and
drop out of the problem.
To preserve the information encoded in 
(\ref{continuity}) and (\ref{magevol}),
we must impose an auxiliary constraint on 
the mass-flux distribution of the final state in order to connect it with the
initial state and uniquely specify the problem.
The constraint expresses the fact that
material cannot flow across
magnetic flux surfaces in the limit $\sigma\rightarrow\infty$.
We delay
consideration of ohmic dissipation,
where magnetic flux diffuses through the fluid
at short length-scales via (\ref{magevol}),
to a future paper.

We define spherical polar coordinates
$(r,\theta,\phi)$ such that $\theta = 0$ defines
the symmetry axis of the pre-accretion magnetic field.
For an axisymmetric configuration,
there exists a scalar flux function $\psi(r,\theta)$
that generates $\vv{B}$ via
\begin{equation}
\vv{B} = \frac{\nabla\psi}{r\sin\theta}\times\hat{\vv{e}}_{\phi}.
\label{Bdef}
\end{equation}
The toroidal
component $B_{\phi}$ is zero at all times, if the accretion process is
axisymmetric  and
$B_{\phi} = 0$ in the initial accretion state.
Upon substituting (\ref{Bdef}) into (\ref{mhdforcestatic}), we obtain
\begin{equation}
\nabla p + \rho\nabla\phi +  (\Delta^2\psi)\nabla\psi = 0\, ,
\label{mhdforcestaticpsi}
\end{equation}
with
\begin{equation}
\Delta^2 =  \frac{1}{\mu_0 r^2\sin^2\theta} \\
\left[ \frac{\partial^2}{\partial r^2} + \\
\frac{\sin\theta}{r^2}\frac{\partial}{\partial\theta}\left(\frac{1}{\sin\theta}\frac{\partial}{\partial\theta}\right)
\right]\, .
\label{gs}
\end{equation}
We can then resolve (\ref{mhdforcestaticpsi}) into components
parallel and perpendicular to
the magnetic field:
\begin{equation}
\rho\nabla\phi + \nabla p = 0 \, ,
\label{alongB}
\end{equation}
\begin{equation}
\rho\nabla\phi + \nabla p + (\Delta^2\psi)\nabla\psi = 0 \, .
\end{equation}

In this paper, we assume the accreted material forms an
isothermal atmosphere, with $p = c_{\rm s}^2 \rho$,
where $c_{\rm s}$ denotes the isothermal sound speed.
[The force equation for a general equation of state
$p = p(\rho)$ is given in Appendix A of
\citet{mou74}.]
The gravitational potential $\phi$, determined
by Poisson's equation, $\nabla^2\phi = 4\pi G \rho$,
is the sum of contributions from the accreted material ($M_{\rm a}$) and the
underlying neutron star ($M_{*}$),
with $M_{\rm a}\ll M_{*}$.
As the hydromagnetic
length-scale $|\vv{B}|/|\nabla\vv{B}|$ is much smaller than
the hydrostatic length-scale $|p|/|\rho\nabla\phi|$
(verified a posteriori), 
$\nabla\phi$ is approximately constant near the stellar
surface for our purposes, i.e.
\begin{equation}
\phi = GM_{*}r/R_{*}^2\, ,
\end{equation}
where $M_{*}$ and $R_{*}$ are the mass and
radius of the neutron star.

We use the method of characteristics 
to solve (\ref{mhdforcestaticpsi})
assuming the gravitational field is radial
($M_{\rm a} \ll M_{*}$).
The $r$ component reads
$\rho_r + (\Delta^2\psi)/c_{\rm s}^2 \psi_{r} = -\rho/c_{\rm s}^2 \phi_r$
and the $\theta$ component reads
$\rho_{\theta} = (\Delta^2\psi)/c_{\rm s}^2 \psi_{\theta}$,
where subscripts indicate differentiation.
Together these become:
$\rho_r +({\psi_r}/{\psi_{\theta}})\rho_{\theta}= -{\rho}/{c_{\rm s}^2} \phi_r$.
The characteristic equation is:
${dr} = ({\psi_{\theta}}/{\psi_r})d\theta = -d\rho/({\phi}{\rho})$.
This is solved to yield the two characteristic curves:
$\log\rho + \phi/c_{\rm s}^2 = C_1$ and $\psi = C_2$.
Thus the characteristic solution is
$\log\rho + \phi/c_{\rm s}^2 = f(\psi)$ or, equivalently,
\begin{equation}
p = F(\psi) \exp[{-(\phi-\phi_0)/c_{\rm s}^2}]
\label{rhoF}
\end{equation}
where $F(\psi) = \exp[f(\psi)]$ is an arbitrary positive function
to be specified and 
$\phi_0 = GM_{*}/R_{*}$ is a reference potential.
This is just the usual barometric formula with a different base
pressure $F(\psi)$ for each field line.
Note that $\nabla F$ is parallel to $\nabla\psi$,
so $\psi$ and hence $F$ are constant
along a field line, and one has
$\nabla F = F^{\prime}(\psi)\nabla\psi$.
Substituting into (\ref{mhdforcestaticpsi}), we obtain a second order,
non-linear, elliptic partial differential equation,
the Grad-Shafranov equation, for $\psi$:
\begin{equation}
\Delta^2\psi = -F^{\prime}(\psi)\exp[-(\phi-\phi_0)/c_{\rm s}^2]\, .
\label{mhdstaticF}
\end{equation}

Equation (\ref{mhdstaticF}) can be understood as follows.
The quantity $F$ is a function of $r$ and $\theta$ through $\psi$
at hydrostatic equilibrium, expressing the fact that
magnetic forces act only
perpendicular to field lines, while pressure gradients balance
gravity along field lines.
If
$\psi(r,\theta)$
is given, and if matter is distributed between
field lines so that the forces parallel to field lines are in exact
balance, then forces perpendicular to the field lines are
brought into
balance by the appropriate current density
$\mu_0^{-1}\nabla\times\vv{B}$.

\subsection{Magnetic flux freezing and mass-flux ratio}
\label{sec:fluxfreeze}

Many authors guess $F(\psi)$ when 
modelling various systems, e.g. structures in the solar corona,
like prominences and arcades
\citep{dun53,low80},
and accreting compact objects
\citep{ham83,uch81,bro98,mel01}.
As $F(\psi)$ does not change in passing from the initial 
(pre-accretion) to the final (post-accretion) state
in the ideal-MHD limit,
the guessed $F(\psi)$
conflicts with the initial $F(\psi)$ except under
very special circumstances.
In this paper, we adopt a self-consistent approach
whereby we calculate $F(\psi)$ explicitly by demanding
that the mass-flux distribution of the final state equals that
of the initial state, plus the accreted material,
as described below.

Define a local coordinate system $(s,t)$, with unit vectors
$\hat{\vv{e}}_s = \vv{B}/|\vv{B}|$ and
$\hat{\vv{e}}_t = \nabla\psi/|\nabla\psi|$
parallel and perpendicular to the magnetic field respectively.
The amount of matter between
two infinitesimally separated flux surfaces $\psi$
and $\psi + d\psi$ is
$dM = 2\pi\int_C ds dt\rho r\sin\theta$,
where
$ds dt =|\hat{\vv{e}}_s ds\times\hat{\vv{e}}_t dt|= ds {d\psi}/{|\nabla \psi|}$
is an infinitesimal area element, and
$C$ is the curve $\psi[r(s),\theta(s)] = \psi$.  Hence we can write 
\begin{equation}
\frac{dM}{d\psi} = 2\pi\int_C ds\rho[r(s), \theta(s)] r\sin\theta|\nabla \psi|^{-1},
\label{dm}
\end{equation}
where $dM/d\psi$ is the mass-flux ratio.
Upon substituting (\ref{rhoF}) into (\ref{dm}), we arrive at
\begin{equation}
\begin{split}
F(\psi)
&=
\frac{c_{\rm s}^{2}}{2\pi}\frac{dM}{d\psi} \\ 
&\times  \left\{\int_{C} ds\, r\sin\theta {|\nabla\psi|}^{-1}e^{-(\phi-\phi_0)/c_{\rm s}^2} \right\}^{-1},
\label{fpsi}
\end{split}
\end{equation}
which is to be solved simultaneously with (\ref{mhdstaticF})
for $\psi(r,\theta)$ given $dM/d\psi$.
For problems involving accretion,
the mass-flux constraint is not conservative;
$dM/d\psi$ in the final state equals $dM/d\psi$
in the initial state plus the mass-flux distribution
of the accreted material.

In disc-fed accretion, mass accretes onto polar magnetic field lines
that close beyond the inner edge of the accretion disc,
located at a radius
\begin{equation}
\label{eq:alfven}
\frac{R_{\rm a}}{R_{*}} \approx \
270 
\left(\frac{\dot{M}}{10^{-9} \, \Msun\rm{yr}^{-1}}\right)^{-2/7}\left(\frac{|\vv{m}|}{10^{20}\rm{T\, m}^3}\right)^{4/7}.
\end{equation}
in the equatorial plane
\citep{gho79,bas76}.
The flux surface that closes at $R_{\rm a}$ is related
to the flux surface at the stellar equator,
$\psi_{*} = \psi(R_{*},\pi/2)$, by
$\psi_{\rm a} = \psi_{*}R_{*}/R_{\rm a}$,
for a dipole.
We do not model the mechanism by which plasma enters the
magnetosphere (e.g.\ Rayleigh-Taylor and Kelvin-Helmholtz
instabilities at $R \approx R_{\rm a}$),
which sets the form of $dM/d\psi$ in reality,
as this is an unsolved problem
\citep{bas76,aro84}.
Instead, we assume that the accreted mass is distributed
nearly uniformly within the polar flux tube $0\leq\psi\leq\psi_{\rm a}$,
and that leakage onto flux surfaces $\psi_{\rm a}\leq\psi\leq\psi_{*}$
is minimal.
A step change in $dM/d\psi$
at $\psi_{\rm a}$ leads to numerical problems,
so we approximate the mass distribution over one hemisphere by
$M(\psi) = M_{\rm a}(1 - e^{-\psi/\psi_{\rm a}})/2(1 - e^{-\psi_{*}/\psi_{\rm a}})\, .$
We have checked that the solution of 
(\ref{mhdstaticF}) and (\ref{fpsi}) is not sensitive to
the exact functional form of $dM/d\psi$.
Finally, we assume that the accreted material does not transport
any magnetic flux, e.g.\ from the accretion disc
(cf. \citealt{uch81}).
If the magnetic dipole moment is less than
$\approx 10^{16}$ T\, m$^{3}$, we have $\psi_{\rm a}\approx\psi_{*}$
\citep{che98} and the above functional form of $M(\psi)$ is
inadequate.
This occurs at the latest stages of accretion
($M_{\rm a} > 0.1\Msun$),
outside the regime modelled in this paper.

\subsection{Initial and boundary conditions}
\label{bcsection}

In this paper, we investigate the distortion of an initially
dipolar magnetic field,
\begin{equation}
\psi_{\rm i}(r,\theta) = \psi_{*}R_{*}r^{-1}\sin^2\theta,
\label{dipole}
\end{equation}
with $\psi_{*} = B_{*}R_{*}^2/2$, where $B_{*}$ is the polar
magnetic field before accretion.
Given $dM/d\psi$ as a function of $M_{\rm a}$,
we solve (\ref{mhdstaticF}) and (\ref{fpsi}) subject to the
Dirichlet boundary conditions
\begin{equation}
\psi(R_{*},\theta) = \psi_{*}\sin^2\theta \quad {\rm and} \quad \underset{r\rightarrow\infty}{\lim}\psi(r,\theta) = 0.
\label{boundary}
\end{equation}
At the surface,
i.e.\ the crystalline layers
of density $\lesssim 4\times 10^{14}$ kg m$^{-3}$,
the field is dipolar.
Far from the star, one has $\psi\propto r^{-1}$, as
for any localised, static current distribution.
The approximation that the surface field remains dipolar at all
times is valid provided that $M_{\rm a}$ is small compared
to $M_{*}$, for then the footpoints of the magnetic field
lines are anchored to the highly conducting, high-inertia
interior of the star.
The surface field may be generated deep in the neutron star
core or by a dynamo in the inner crust
\citep{tho93,kon01}.
This line-tying boundary condition is a feature of models of
magnetic loops in the solar corona \citep{low80,zwe82}
and earlier work on neutron star accretion
\citep{uch81,ham83,bro98,lit01}.
A drawback of preventing the accreted matter from sinking is that
unrealistically high densities ($\geq 4\times 10^{14}$ kg m$^{-3}$)
are produced locally, at the base of the column, for
$M_{\rm a} > 10^{-8}\Msun$.
Recent modelling of the magnetic field beneath the surface
of an accreting neutron
star, based on (\ref{magevol}) with a velocity distribution
for the superfluid assumed,
illustrates the effects of 
submergence and subsequent incorporation of accreted matter
into the crust
\citep{cho02}.
\citet{cum02} discusses a \emph{vacuum} (like in this paper)
 and a \emph{screened} boundary condition
at the surface.

In the numerical calculations presented in Section \ref{results},
two extra grid-related boundaries are introduced:
the outer radius of the grid, at $r = R_{\rm m}$, and
the lines $\theta = 0$ and $\theta = \pm\pi/2$, arising when 
the grid is restricted to one quadrant
or hemisphere.
We choose $R_{\rm m}$ large enough to include the layer where the
accretion-induced screening currents lie, i.e., above the greater
of the hydrostatic ($c_{\rm s}^2/g$) and Alfv\'en
($|\vv{B}|/|\nabla\vv{B}|$) scale heights.
In practice, this is achieved by increasing $R_{\rm m}$ until the
dipole moment of the solution varies by less than $0.1\%$.
At $r = R_{\rm m}$, the magnetic field is taken to be radial,
with $\partial\psi/\partial r(R_{\rm m},\theta) = 0$,
i.e.\ a free boundary.  (It is outside the scope of this
paper to model the disc-magnetosphere interface in detail;
(see \citealt{ras99a}).
Another possible way to treat the boundary condition at $r = R_{\rm m}$
is to set $\psi(R_{\rm m},\theta) = \psi_{\rm m}\sin^{2}\theta$
and adjust $\psi_{\rm m}$ iteratively to give the
self-consistent dipole moment of the solution, but
we encountered numerical difficulties with this approach.

There are two physically plausible choices for the polar
and equatorial boundary conditions:
(i) $\psi(r,\pi) = 0$ and 
$\partial\psi/\partial\theta(r,\pi/2) = 0$,
as for the initial dipole, or
(ii)
$\partial\psi/\partial\theta(r,\pm\pi/2) = 0$,
north-south symmetry.
We mostly adopt (i)
but explore (ii) for completeness in Section \ref{sec:polarcap},
where it is shown that field lines on either side
of the pole are peeled
away, leaving the $\psi = 0$ line isolated,
without affecting the dipole moment significantly.
Strictly speaking, the conditions
$\partial\psi/\partial\theta(r,\pi/2) = 0$
and
$\partial\psi/\partial r(R_{\rm m},\pi/2) = 0$
force the magnetic field to vanish artificially at
$(R_{\rm m},\pi/2)$,
but $|\vv{m}|$ is affected by less than 0.1\%
(see Section \ref{results}).

\section{SOLUTION METHODS}
\label{sec:solutionmethods}

In this section, we discuss three ways to solve
(\ref{mhdstaticF}) and (\ref{fpsi}):
analytically by Green functions (Section \ref{greenfunction}),
analytically in the small-$M_{\rm a}$ approximation
(Section \ref{smallma}),
and numerically, by an iterative algorithm due to
\citet{mou74}
(Sections \ref{iterativescheme} and \ref{sec:convergence}).

\subsection{Green functions}
\label{greenfunction}

The Grad-Shafranov boundary value problem
(\ref{mhdstaticF}),
\begin{equation}
\Delta^2\psi(r,\theta) = Q(r,\theta),
\end{equation}
with
\begin{equation}
\psi(R_{*},\theta) = \psi_{*}\sin^2\theta \quad {\rm and} \quad \
\lim_{r\rightarrow\infty}\psi(r,\theta) = 0,
\end{equation}
can be solved analytically by Green functions if the source term
$Q(r,\theta)$ is known as a function of $r$ and $\theta$.
In principle,
$Q(r,\theta)$ is given by (\ref{mhdstaticF}) and
(\ref{fpsi}); in practice, it 
is not known analytically.
With $\psi$ specified on the boundary $S$ of the volume $V$,
we can write
(see Appendix \ref{greenappendix1})
\begin{equation}
\begin{split}
\label{psisolution}
\psi(\vv{x}) = & \int_V d^3\vv{x}^{\prime}\, G^{*} Q\,\, \\&
+\int_S d^2\vv{x}^{\prime}\, (\psi\nabla G^{*}-G^{*}\nabla\psi+\vv{b}\psi G^{*}),
\end{split}
\end{equation}
where $G$ and $G^{*}$ are Green functions for
$L = \mu_0 r^2\sin^2\theta\nabla^2$ and its adjoint $L^{*}$,
satisfying
\begin{equation}
\label{gsmuG}
\frac{\partial^2G}{\partial r^2} + \\
\frac{(1-\mu^2)}{r^2}\frac{\partial^2G}{\partial\mu^2} \\
= \frac{1}{r^2}\delta(r - r^{\prime})\delta(\mu - \mu^{\prime})\,~,
\end{equation}
and
\begin{equation}
\begin{split}
\label{gsmuGs}
\frac{\partial^2G^{*}}{\partial r^2} + \
\frac{(1-\mu^2)}{r^2}\frac{\partial^2G^{*}}{\partial\mu^2} \
+ \frac{4}{r}\frac{\partial G^{*}}{\partial r} \
- \frac{4\mu}{r^2}\frac{\partial G^{*}}{\partial\mu} \\
 = \frac{1}{r^2}\delta(r - r^{\prime})\delta(\mu - \mu^{\prime})\,~,
\end{split}
\end{equation}
with
$\mu = \cos\theta$ and
$\vv{b} = -2r^{-1}(\hat{\vv{e}}_{r} + \cot\theta\hat{\vv{e}}_{\theta})$.
Upon solving (\ref{gsmuG}) and (\ref{gsmuGs}), we obtain
\begin{equation}
\begin{split}
\label{gs1}
G(r,\mu, r^{\prime},\mu^{\prime})
 &=
\sum_{\ell = 0}^{\infty} N_{\ell}^{-1} g_{\ell+1}(r, r^{\prime}) \\
 &\times (1-\mu^2)C_{\ell}^{3/2}(\mu^{\prime})C_{\ell}^{3/2}(\mu)\,~,
\end{split}
\end{equation}
and
\begin{equation}
\begin{split}
\label{gs2}
G^{*}(r,\mu, r^{\prime},\mu^{\prime})
 &= \sum_{\ell = 0}^{\infty} N_{\ell}^{-1} g^{*}_{\ell}(r, r^{\prime}) \\
 &\times (1-\mu^{\prime 2})C_{\ell}^{3/2}(\mu^{\prime})C_{\ell}^{3/2}(\mu)\,~,
\end{split}
\end{equation}
with
\begin{equation}
g_{\ell}(r, r^{\prime}) = \frac{1}{(2\ell+1)r^{\prime 2}} \frac{r_{<}^{\ell+1}}{r_{>}^{\ell}}\left[ \left(\frac{R_{*}}{r_{<}}\right)^{2\ell+1} -1\right],
\end{equation}
$r_{<} = \min(r,r^{\prime})$,
$r_{>} = \max(r,r^{\prime})$,
\begin{equation}
 g^{*}_{\ell}(r, r^{\prime}) = \left(\frac{r^{\prime}}{r}\right)^{2}g_{\ell+1}(r, r^{\prime}),
\end{equation}
\begin{equation}
N_{\ell} = 2(\ell +1)(\ell +2)(2\ell +3)^{-1}~,
\end{equation}
and hence,
from (\ref{psisolution}), we arrive at the complete solution
\begin{equation}
\begin{split}
\label{completesolution}
\psi(r,\mu)
  =
 \frac{\psi_{*}R_{*}(1-\mu^2)}{r}\,  + 
 \newline
 (1-\mu^2)\sum_{\ell = 0}^{\infty}N_{\ell}^{-1}C_{\ell}^{3/2}(\mu) \\
\times 
\int_{-1}^{1}d\mu^{\prime}\int_{R_{*}}^{\infty}dr^{\prime} r^{\prime 2}
g_{\ell}^{*}(r^{\prime},r)C_{\ell}^{3/2}(\mu^{\prime})Q(r^{\prime},\mu^{\prime}). \\
\end{split}
\end{equation}
$C_{\ell}^{3/2}(\mu)$ denotes a Gegenbauer polynomial of order $\ell$
(see Appendix \ref{greenappendix}).

\subsection{Analytic approximation for small $M_{\rm a}$ }
\label{smallma}

In the limit of 
small $M_{\rm a}$, where $dM/d\psi$ and hence
$Q(r,\theta)$
are small,
one can show (see Appendix \ref{greenappendix3})
that the magnetic flux distribution reduces to
\begin{equation}
\psi(r,\theta) = \psi_{\rm i}(r,\theta)(1 - b^2 M_{\rm a}/M_{\rm c})
\end{equation}
far from the star ($r \rightarrow \infty$), 
with
\begin{equation}
M_{\rm c} = 2 \pi G M_{*}\psi_{*}^2/\mu_{0}c_{\rm s}^4R_{*}^{2}
\end{equation}
For convenience, we write this using CGS units
\begin{equation}
\label{eq:mc}
\frac{M_{\rm c}}{\Msun} = \
1.2 \times 10^{-4} \
\left(\frac{c_{\rm s}}{10^{8} \, {\rm cm \, s^{-1}}}\right)^{-4}\left(\frac{B_{*}}{10^{12} {\rm G}}\right)^{2}.
\end{equation}
where
$M_{*} = 1.4 \Msun$ and
$R_{*} = 10^{6} {\rm cm}$.

It follows that the magnetic dipole moment scales as
$|\vv{m}| = |\vv{m}_{\rm i}|(1 - M_{\rm a}/M_{\rm c})$.
This scaling agrees, in the small-$M_{\rm a}$ limit,
with empirical scalings of the form
$|\vv{m}| = |\vv{m}_{\rm i}|(1 + M_{\rm a}/M_{\rm c})^{-1}$,
with
$M_{\rm c}\approx 10^{-5}\Msun$,
that have been proposed in the literature
\citep{shi89,che98}.

\subsection{Iterative numerical scheme}
\label{iterativescheme}

To solve
(\ref{mhdstaticF}) and (\ref{fpsi})
self-consistently for $\psi(r,\theta)$ for 
$M_{\rm a} \gg M_{\rm c}$,
we employ an iterative numerical algorithm similar to the one
introduced by \citet{mou74} to study the Parker instability
of the Galactic magnetic field.
The algorithm and its performance are discussed in detail
in Appendices \ref{greenappendix} and \ref{iterationappendix}
and summarized briefly here.

Given $dM/d\psi$ and an initial guess
$\psi^{(0)}(r,\theta)$,
we calculate the locations of $N_c$ contours of $\psi$,
spaced either linearly or logarithmically in $\psi$,
capturing topologically disconnected contours and
closed loops
\citep{sny78}.
We then compute $F[\psi^{(0)}]$ from (\ref{fpsi})
and hence $F^{\prime}[\psi^{(0)}]$ after polynomial fitting
(simple differencing causes numerical difficulties;
see Appendix \ref{iterationappendix3}).
The Poisson equation (\ref{mhdstaticF}) is solved with this
source term using
successive overrelaxation to obtain $\psi_{\rm new}^{(0)}(r,\theta)$,
and the next iterate is obtained by underrelaxation:
$\psi^{(n+1)} = \Theta^{(n)}\psi^{(n)} + [1 - \Theta^{(n)}]\psi_{\rm new}^{(n)}$,
with $0\leq\Theta^{(n)}\leq 1$.
Iteration continues until the convergence criterion
${|\psi_{\rm new}^{(n+1)}-\psi^{(n)}|} <  \epsilon {|\psi_{\rm new}^{(n+1)}|}$
is satisfied on average across the grid.  We usually take
$\epsilon = 10^{-2}$ in this paper.
Physically, the algorithm starts from a
trial magnetic field (and associated current distribution),
the accreted
mass $M_{\rm a}$ is distributed among the flux tubes
according to $dM/d\psi$,
mass is allowed to slide up or down flux tubes
to achieve hydrostatic equilibrium along $\vv{B}$,
a new current distribution is computed that balances
forces perpendicular to $\vv{B}$, and
the process is repeated.

\subsection{Numerical convergence}
\label{sec:convergence}

There is no general rule for choosing $\Theta^{(n)}$.
We find, by experimentation, that one must decrease
$1-\Theta^{(n)}$ as $M_{\rm a}$ increases;
a useful rule of thumb is
$1-\Theta^{(n)} \approx
(M_{\rm a}/10^{-7}\Msun)^{-1}(\psi_{*}/10\psi_{\rm a})^{-2}$,
for $M_{\rm a}\geq 10^{-7}\Msun$.
More details can be found in Table \ref{table:varyTheta} and
Appendix \ref{sec:underrelax}.
At least $2/[(1-\Theta)\log_{10}(\epsilon)]$
iterations are required for
convergence;
bootstrapping is recommended,
i.e.\ using the
equilibrium solution for a lower value of $M_{\rm a}$
as the first iterate instead of
the dipole.
We show in Appendix \ref{iterationappendix5},
that the error in $\psi$ averaged over the grid, scales as
$G^{-1.6}$,
where $G$ is the number of grid cells in each dimension..
The optimum number of contours is
$N_c \approx G - 1$;
$F^{\prime}(\psi)$
becomes jagged for $N_c \gg G$ due to grid crossings,
(as demonstrated in Figure \ref{fig:converge}).
To concentrate maximum grid resolution near the stellar surface 
and at the edge of the polar cap ($\psi = \psi_{\rm a}$),
where screening currents predominantly reside and gradients of
$\rho$ and $\psi$ are steepest, we scale the $r$ and $\theta$
coordinates logarithmically as described in
Appendix \ref{app:logscale}.

Figure \ref{fig:resid} displays the mean residual as a function of 
iteration number.
Convergence is rapid for $M_{\rm a}\leq 10^{-6}\Msun$
and poor for $M_{\rm a}\geq 10^{-4}\Msun$.
Large fluctuations in the mean residual are mainly due to
the polynomial fit to $F(\psi)$ and
the appearance of magnetic bubbles
(see Section \ref{sec:bubbles}).
\begin{center}
\begin{figure}
\centering
\includegraphics[height=65mm]{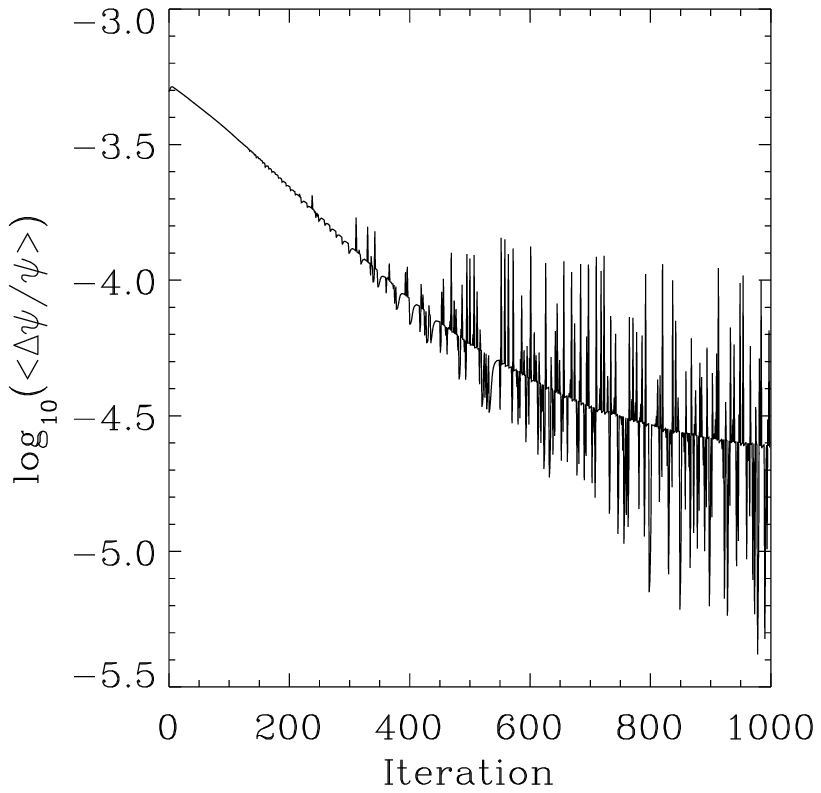} 
\includegraphics[height=65mm]{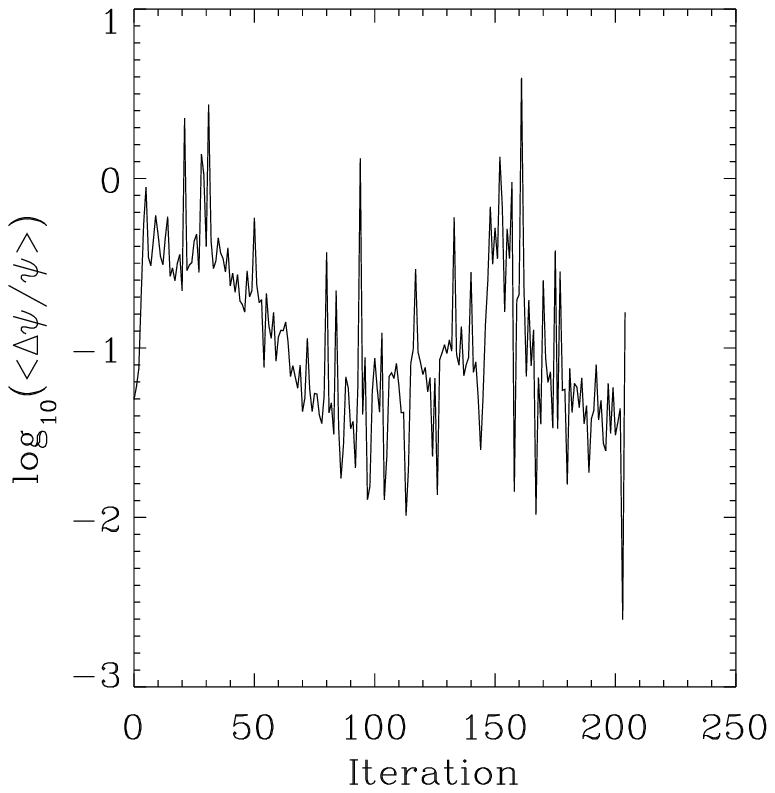} 
\caption{\small
Mean residual versus iteration number.
$M_{\rm a} = 10^{-6}\Msun$,
$\theta = 0.995$ (\emph{top}).
$M_{\rm a} = 10^{-5}\Msun$, $\theta = 0.95$ (\emph{bottom}).
}
\label{fig:resid}
\end{figure}
\end{center}

\section{EQUATORWARD HYDROMAGNETIC SPREADING}
\label{results}

In this section, we present the results of the analytic
and numerical calculations described in Section
\ref{sec:solutionmethods}.
The hydromagnetic structure of the polar `mountain'
formed by the accreted material is described
in Sections \ref{sec:tutu} and \ref{sec:massflux}
and compared with previous calculations in which
$F(\psi)$ is arbitrary
(e.g.\ \citealt{bro98}).
The physics of equatorward spreading of the accreted
material and the formation of an equatorial magnetic
`tutu', including a criterion for the onset of spreading,
is discussed in Section \ref{sec:spreading}.
The scalings of $|\vv{m}|$ with respect to
$M_{\rm a}$ and $b = \psi_{*}/\psi_{\rm a}$
are derived analytically and numerically in
Sections \ref{sec:dipolereduce} and \ref{sec:polarcap}.
Finally, the formation of magnetic bubbles disconnected from the
star --- a new effect --- is explored in
Section \ref{sec:bubbles}.
We start from the undisturbed dipole (\ref{dipole})
and adopt the following physical parameters:
$M_{*} = 1.4\Msun$, $R_{*} = 10^4$m,
$B_{*} = 10^{8}$ T \citep{har97},
$c_{\rm s} = 10^6$m s$^{-1}$,
$x_{0} = c_{\rm s}^2R_{*}^{2}/GM_{*} = 0.54$m and hence
$a = R_{*}/x_{0} = 1.86\times 10^{4}$
\citep{bro98}.
The results are mostly presented in rectangular $(r,\theta)$ plots
scaled logarithmically where
appropriate to emphasize the boundary layer
of compressed magnetic field.
\begin{center}
\begin{figure}
\centering
\includegraphics[height=65mm]{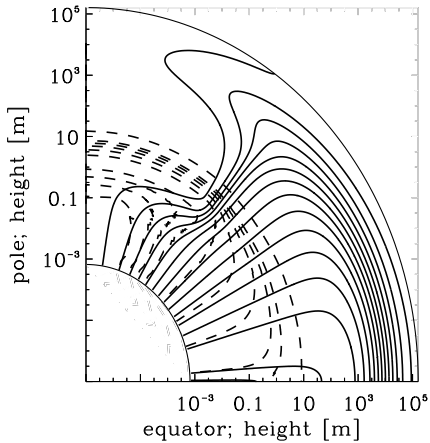} 
\caption{\small
Polar plot of equilibrium magnetic field lines (solid curves)
and density contours (dashed curves) for
$M_{\rm a} = 10^{-5}\Msun$.
The coordinates measure altitude above the stellar surface.
Density contours are drawn for $\eta\rho_{\rm max}$
(maximum at the pole,
$\rho_{\rm max} = 2.52\times 10^{17} {\rm kg m}^{-3}$)
with $\eta = 0.8,\, 0.6,\, 0.4,\, 0.2,\, 0.01,\, 0.001,\, 10^{-4},\,
10^{-5},\, 10^{-6},\, 10^{-12}$.
}
\label{fig:polar}
\end{figure}
\end{center}
\begin{center}
\begin{figure}
\centering
\includegraphics[height=65mm]{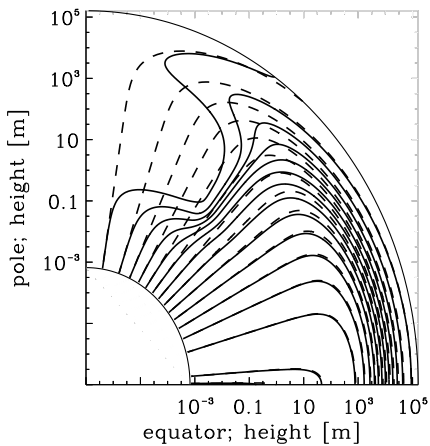} 
\caption{\small
Polar plot of magnetic field lines
after accretion (solid curves)
and before accretion (dashed curves) for
$M_{\rm a} = 10^{-5}\Msun$.
The coordinates measure altitude above the stellar surface.
}
\label{fig:polarinit}
\end{figure}
\end{center}

\subsection{Structure of the polar mountain and equatorial magnetic tutu}
\label{sec:tutu}

During the early stages of accretion, matter piles up on the
polar cap, confined by the tension of the polar magnetic flux tube.
However, for $M_{\rm a} \gtrsim 10^{-5}\Msun$,
the hydrostatic pressure
at the base of the accretion column overcomes
the magnetic tension and 
matter spreads over the stellar surface
towards the equator, dragging along
frozen-in polar field lines.
The spreading distorts 
$\vv{B}$, generating screening currents
$\mu_{0}^{-1}\nabla\times\vv{B}$,
which act to
decrease the magnetic dipole moment
($|\vv{m}|$ is dominated by the polar field).
In its turn, the spreading is also counteracted 
by the tension of the magnetic field lines compressed
towards the equator.
These equatorial stresses, neglected in previous work
\citep{ham83,bro98,lit01}, greatly increase
the $M_{\rm a}$ required to reduce $|\vv{m}|$.

\begin{center}
\begin{figure*}
\begin{tabular}{cc}
 \begin{tabular}{c}
 \includegraphics[height=65mm]{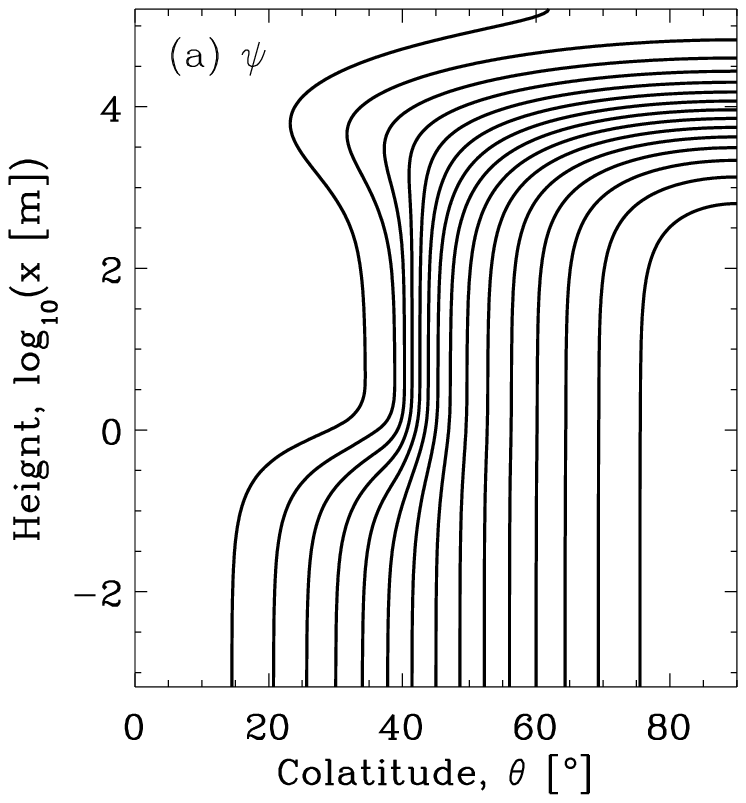} 
 \end{tabular}
 &
 \begin{tabular}{c}
 \includegraphics[height=65mm]{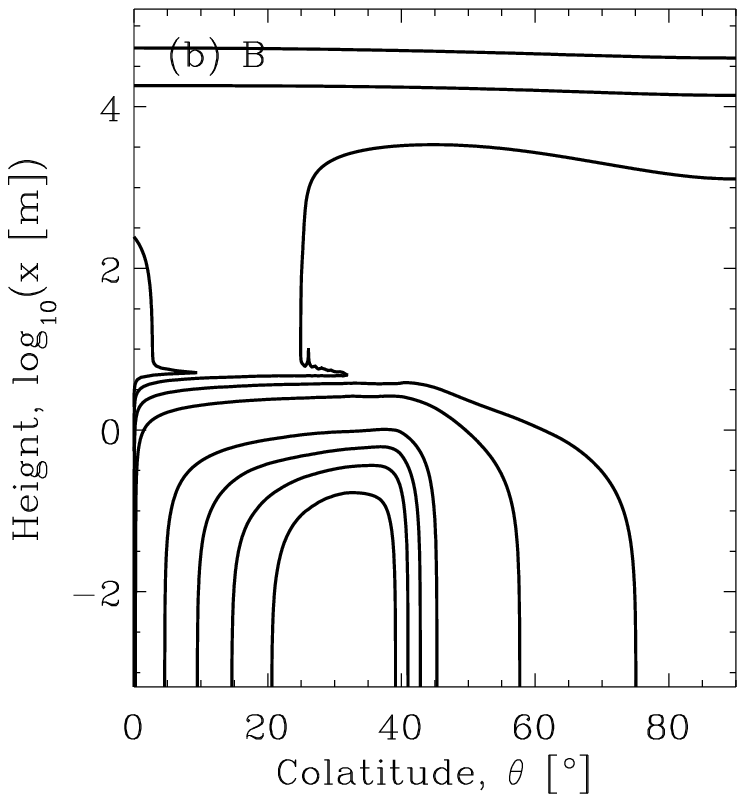} 
 \end{tabular}
 \\
 \begin{tabular}{c}
 \includegraphics[height=65mm]{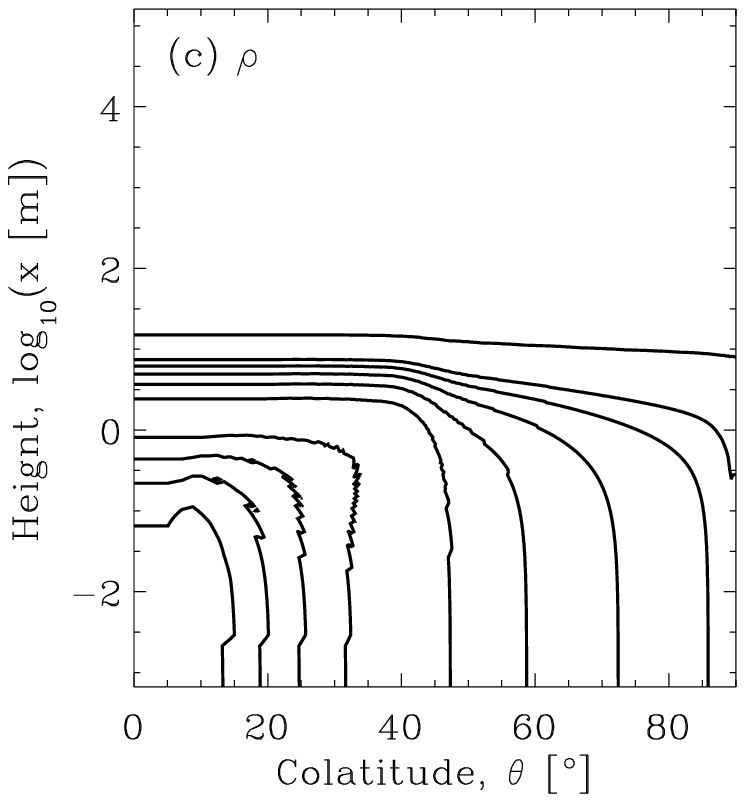} 
 \end{tabular}
 &
 \begin{tabular}{c}
 \includegraphics[height=65mm]{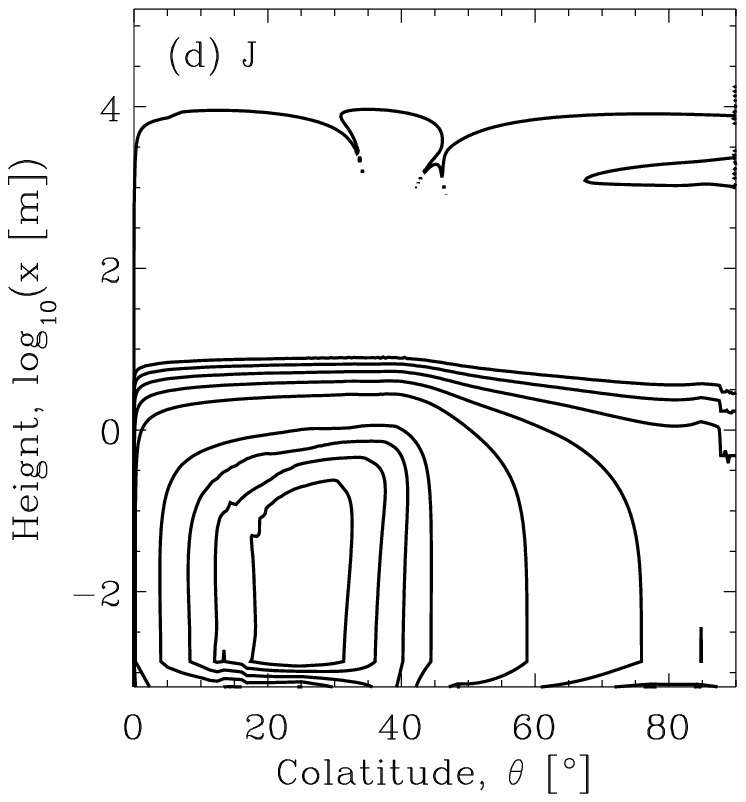} 
 \end{tabular}
 \\
 \begin{tabular}{c}
 \includegraphics[height=65mm]{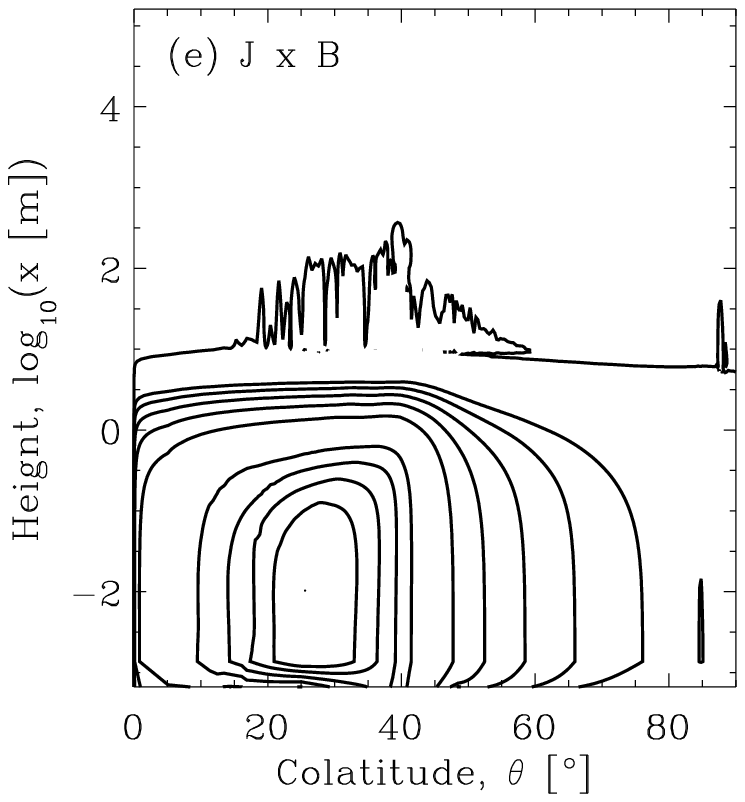} 
 \end{tabular}
 &
 \begin{tabular}{c}
 \includegraphics[height=65mm]{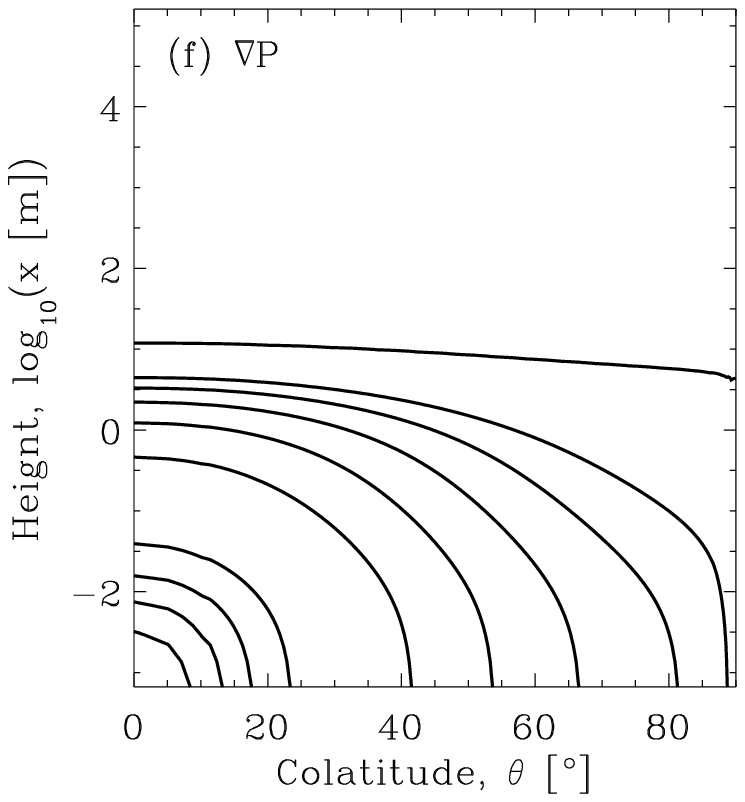} 
 \end{tabular}
\end{tabular}
\caption{\small
$M_{\rm a}=10^{-5}\Msun$.
(a) Magnetic field lines ($\psi$ contours),
(b) magnetic field strength ($|\vv{B}|$ contours),
(c) density,
(d) current,
(e) Lorentz force,
and
(f) pressure gradients.
For each quantity $x$, values $\eta x_{\rm max}$ are plotted, with
$\eta = 0.8,\ 0.6,\ 0.4,\ 0.2,\ 0.01,\ 0.001,\
10^{-4},\ 10^{-5},\ 10^{-6},\ 10^{-12}$.
Maximum values are found to be
$\rho_{\rm max} = 1.7\times 10^{17} {\rm kg m}^{-3}$,
$|\vv{B}|_{\rm max} = 3.9\times 10^{11} {\rm T}$,
$|\vv{J}|_{\rm max} = 2.0\times 10^{15} {\rm A m}^{-2}$,
$|\vv{J}\times\vv{B}|_{\rm max} = 3.3\times 10^{24} {\rm N m}^{-3}$,
$|\nabla P|_{\rm max} = 1.9\times 10^{28} {\rm N m}^{-3}$.
}
\label{fig:Ma5}
\end{figure*}
\end{center}

Figure \ref{fig:polar} shows the magnetic field configuration
and density profile for
$M_{\rm a} = 10^{-5}\Msun$ in cross-section
(cf. schematic version in Figure 1 of \citealt{mel01}).
The `polar mountain' of accreted material is readily
apparent,
traced out by the dashed contours.
Figure \ref{fig:polarinit} shows the distorted magnetic
field configuration overlaid on the field lines of the
undisturbed dipole.
The distorted field exhibits a
pinched, flaring geometry,
termed an `equatorial tutu' by \citet{mel01}.
A more complete view of the overall hydromagnetic
structure can be gained from
Figure \ref{fig:Ma5}.
The tutu-like field is shown again in Figures
\ref{fig:Ma5}(a) and \ref{fig:Ma5}(b),
while the polar mountain ($\rho$) is shown in
Figure \ref{fig:Ma5}(c).
In Figure \ref{fig:Ma5}(d), where the radius of curvature of
$\vv{B}$ is smaller than the hydrostatic scale height $x_0$,
the toroidal screening currents are confined below
altitude $x_0$,
and are concentrated near the polar cap.
(Note that ohmic dissipation, neglected here,
is important at these scales).
The $\vv{J}\times\vv{B}$ force per unit volume
[Figure \ref{fig:Ma5}(e)]
balances the pressure gradient
[Figure \ref{fig:Ma5}(f)],
preventing the accreted material from spreading
all the way to the equator.

The maximum density, attained at
$(r,\theta)=(R_{*},0)$,
is found empirically to be
$\rho_{\rm max} = M_{\rm a}b^{2}/(2x_{0}^{3}\pi a^2)\approx 6\times 10^{12}(b/10)^{2}(M_{\rm a}/10^{-10}\Msun)\,{\rm kg\, m}^{-3}$,
in accord with analytic estimates for
$\rho(r,\theta) \approx \rho_{\rm max}\exp(-x/x_0)\exp(-\psi/\psi_{\rm a})$
carried out in Appendix \ref{greenappendix}.
Consequently, $\rho_{\rm max}$ exceeds the crustal density
$4\times 10^{14} {\rm \, kg\, m}^{-3}$
for $M_{\rm a}\geq 10^{-8}\Msun$.
In reality, this overdensity is moderated by sinking
\citep{bro98,cho02},
which is prevented by the hard surface in our calculation.
(We can alleviate the overdensity
in our model by relaxing the isothermal assumption or
allowing a nonbarometric density distribution along contours.)
For $M_{\rm a} \geq 10^{-6}\Msun$, the maximum magnetic field
strength becomes unrealistically large ($B_{\rm max} \gtrsim 10^{11}$T)
below an altitude $x_0$,
in response to $\rho_{\rm max}$.
Such field strengths formally exceed
the yield stress of the crust
\citep{rom90}.
\begin{center}
\begin{figure*}
\begin{tabular}{cc}
\includegraphics[height=65mm]{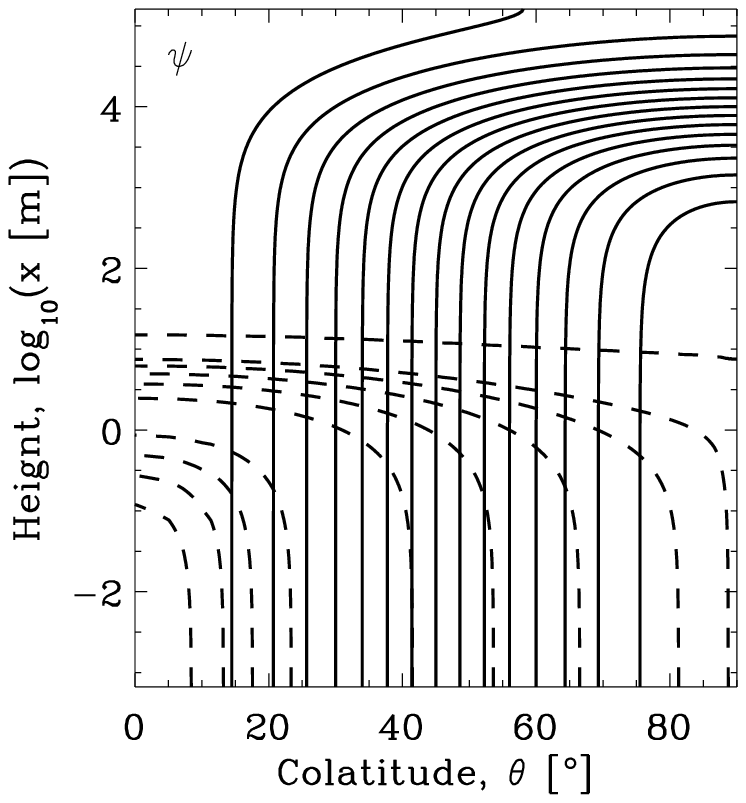} 
&
\includegraphics[height=65mm]{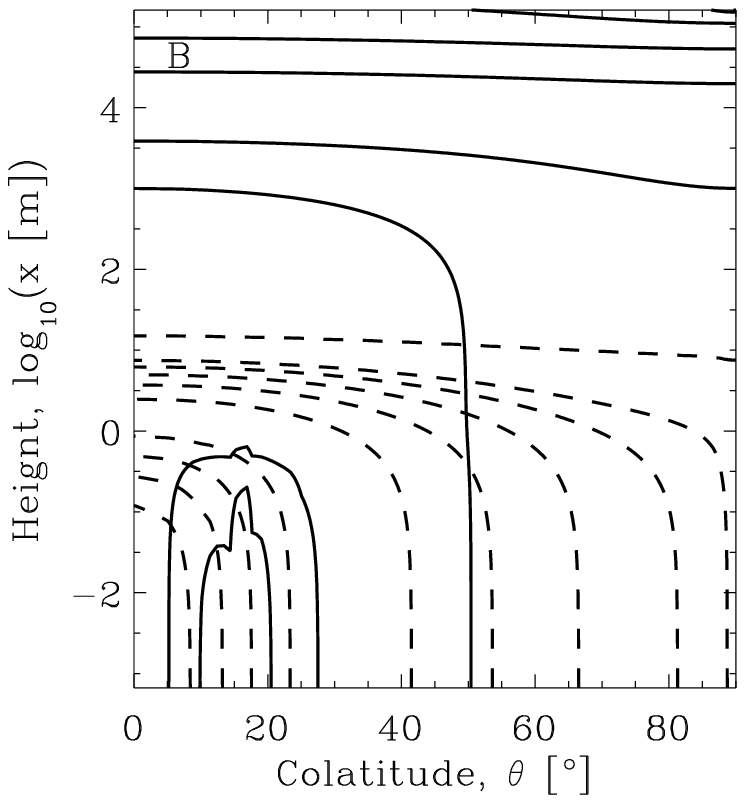} 
\\
\includegraphics[height=65mm]{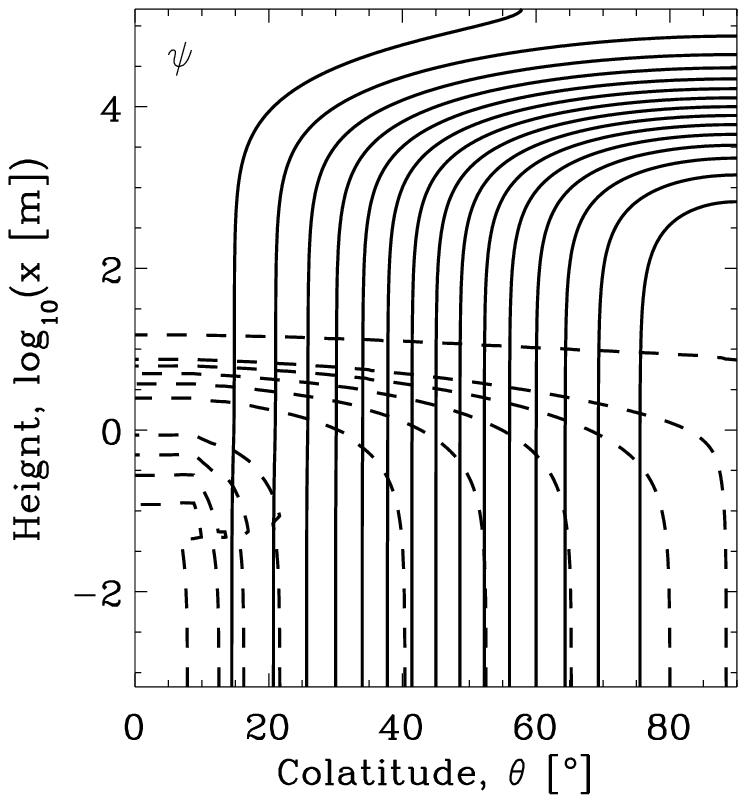} 
&
\includegraphics[height=65mm]{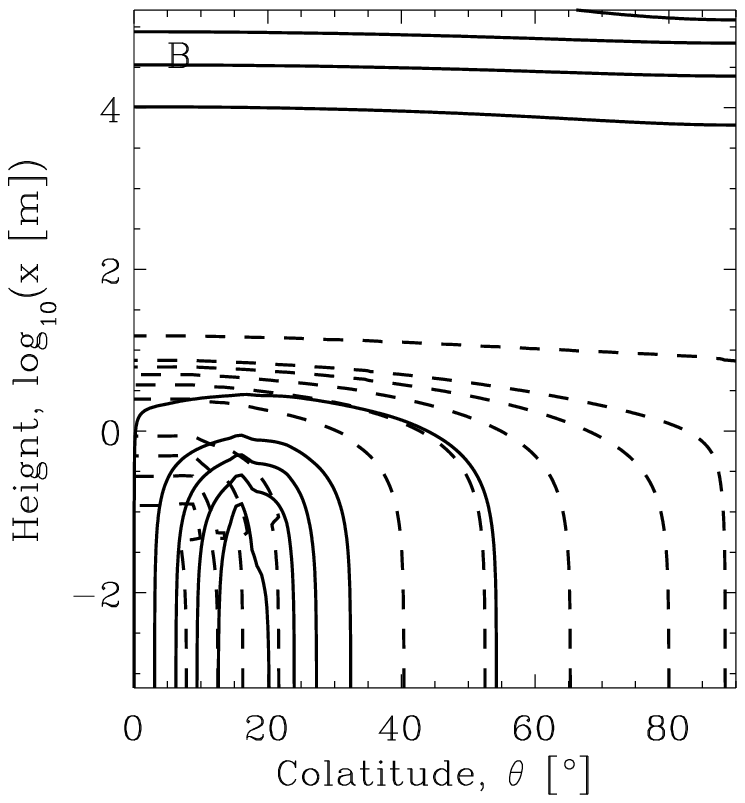} 
\\
\includegraphics[height=65mm]{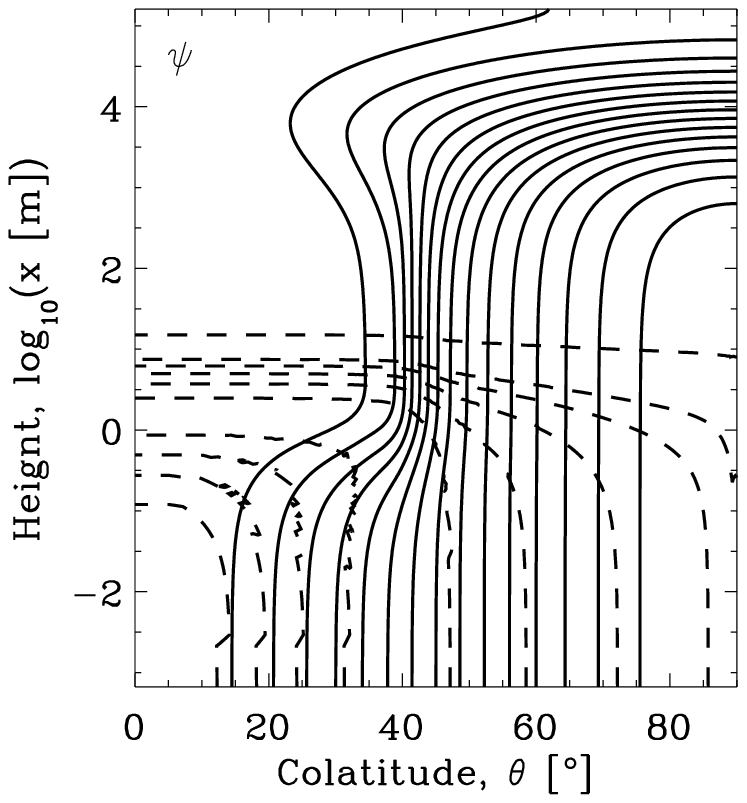} 
&
\includegraphics[height=65mm]{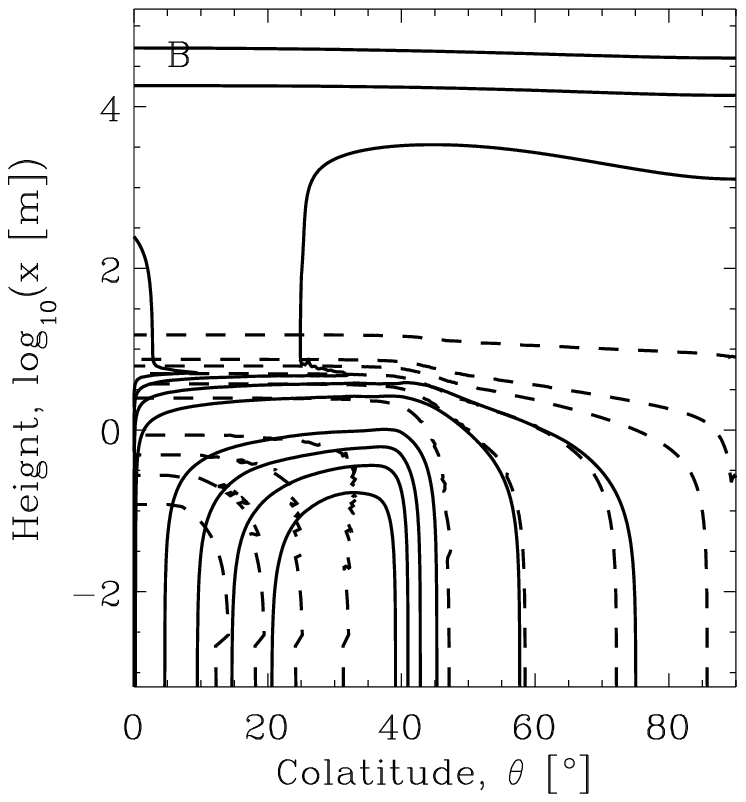} 
\end{tabular}
\caption{\small
Magnetic configuration as a function of accreted mass. 
$M_{\rm a} = 10^{-9}\Msun$,\
$\rho_{\rm max} = 4.2\times 10^{13} {\rm kg m}^{-3}$,\
$|\vv{B}|_{\rm max} = 3.5\times 10^{8} {\rm T}$ (\emph{top});
$M_{\rm a} = 10^{-7}\Msun$,\
$\rho_{\rm max} = 4.2\times 10^{15} {\rm kg m}^{-3}$,\
$|\vv{B}|_{\rm max} = 1.1\times 10^{10} {\rm T}$ (\emph{middle});
and $M_{\rm a} = 10^{-5}\Msun$,\
$\rho_{\rm max} = 1.7\times 10^{17} {\rm kg m}^{-3}$,\
$|\vv{B}|_{\rm max} = 3.9\times 10^{11} {\rm T}$ (\emph{bottom}).
Displayed are contours of $\psi$
(\emph{left, solid}), 
$|\vv{B}|$ (\emph{right, solid}) and $\rho$
(\emph{left and right, dashed}).
with values $\eta\rho_{\rm max}$
and $\eta|\vv{B}|_{\rm max}$, where
$\eta = 0.8,\ 0.6,\ 0.4,\ 0.2,\ 0.01,\ 0.001,\ 10^{-4},\
10^{-5},\ 10^{-6},\ 10^{-12}$.
}
\label{fig:varyMa}
\end{figure*}
\end{center}

\subsection{Sinking}
	The proportion of the accreted material that sinks is
not well constrained.
We consider a crude model of sinking in which a proportion
$s$ of the accreted matter sinks, leaving a proportion $(1-s)$
to spread.  This is modelled by setting
\begin{equation}
d{M}/d{\psi} = (M_{\rm a}/\psi_{\rm a})[(1-s)e^{-\psi/\psi_{\rm a}}/2(1 - e^{-\psi_{*}}) + s/(2b)]\, .
\end{equation}

We find that for $M_{\rm a} = 10^{-5}\Msun$, for s = 0.9
(i.e. all but $10^{-6}\Msun$ redistributed.), the resultant dipole
moment increases from $\approx 0.91$ to $0.96$.

\subsection{Self-consistent mass-flux distribution}
\label{sec:massflux}

\begin{center}
\begin{figure}
    \includegraphics[height=60mm]{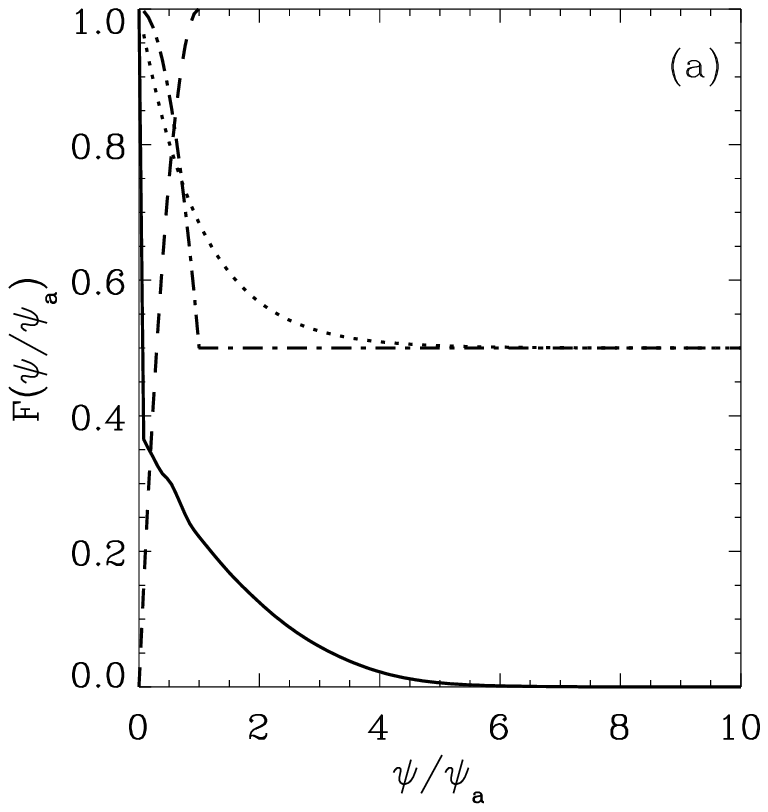}
    \includegraphics[height=60mm]{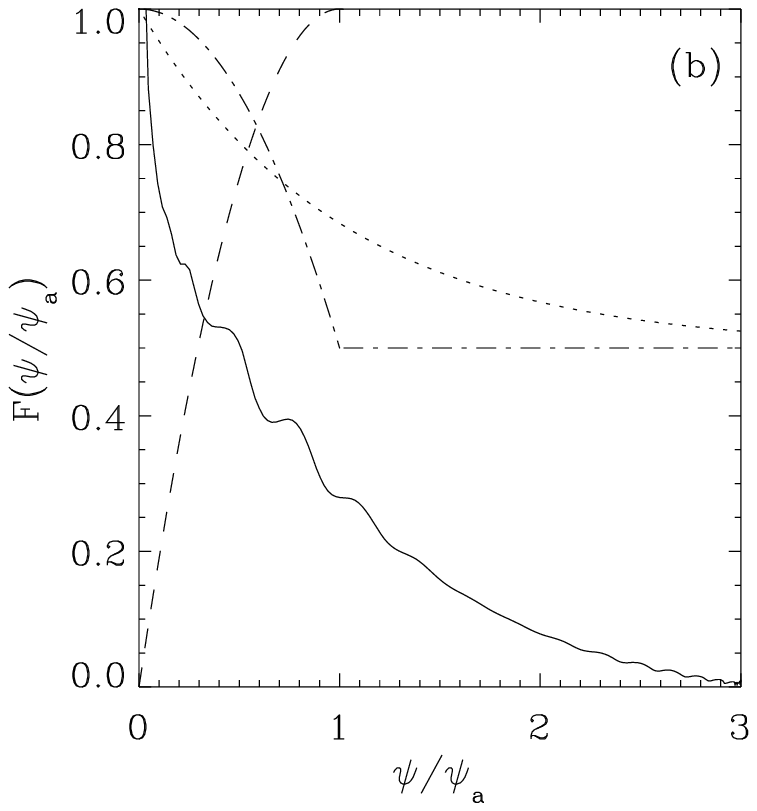}
\caption{\small
Comparison of the self-consistent
$F(\psi)$ (\emph{solid}) for
(a) $M_{\rm a} = 2\times 10^{-5}\Msun, b = 10$ and
(b) $M_{\rm a} = 1.5\times 10^{-4}\Msun, b = 3$
with others in the literature:
\citet{bro98} (\emph{dotted})
\citet{lit01} (\emph{dot-dashed}),
and
\citet{mel01} (\emph{dashed}).
These curves are scaled to the value at the pole, $F(0)$.
}
\label{fig:compareF}
\end{figure}
\end{center}

In previous studies of neutron star accretion,
$F(\psi)$ was chosen arbitrarily
\citep{uch81,ham83,bro98,lit01,mel01}.
In this paper, by contrast, we determine $F(\psi)$
self-consistently by solving
(\ref{mhdstaticF}) and (\ref{fpsi})
simultaneously for a physically plausible choice of
$dM/d\psi$ that places most of the accreted material
at the poles of the initial, undisturbed dipole.
Figures \ref{fig:compareF}(a) and
\ref{fig:compareF}(b)
compare our self-consistent
$F(\psi)$ against the functional forms guessed by previous
authors for two values of $M_{\rm a}$.
The differences are significant,
especially near the pole ($\psi \approx 0$).
[An analagous difference was discovered by \citet{mou74}
when solving for the final states of the Parker instability
in the Galaxy self-consistently,
relative to the previous guess of \citet{par66}.]
A polynomial fit in the numerical code yields the approximate
form
\begin{equation}
\begin{split}
F(\tilde{\psi})
 &=  \exp(\tilde{\psi})(0.027\tilde{\psi}^4-0.13\tilde{\psi}^3 \\
 & \quad + 0.21\tilde{\psi}^2-0.021\tilde{\psi}+0.1333)
\end{split}
\end{equation}
for $N_p = 4, M_{\rm a} = 1.5\times 10^{-4}\Msun, b = 3$,
$\tilde{\psi} = \psi/\psi_{\rm a}$,
and the choice of $dM/d\psi$ in Section \ref{sec:fluxfreeze}.
\begin{center}
\begin{figure}
    \includegraphics[height=60mm]{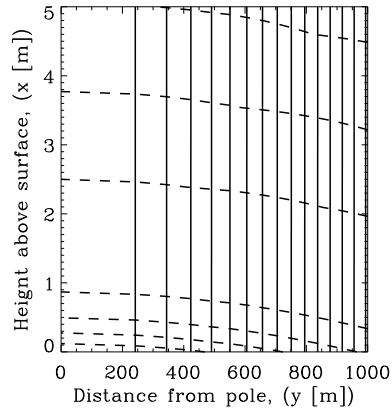}
\caption{\small
Magnetic field lines (\emph{solid}) and density contours (\emph{dashed})
around the polar cap
for $M_{\rm a} = 10^{-13}\Msun$ and $b = 100$; the parameters used in
Figure 2 of \citet{lit01}.
}
\label{fig:comparelitwin}
\end{figure}
\end{center}
\begin{center}
\begin{figure}
 \begin{tabular}{c}
  (a) \\
  \includegraphics[height=60mm]{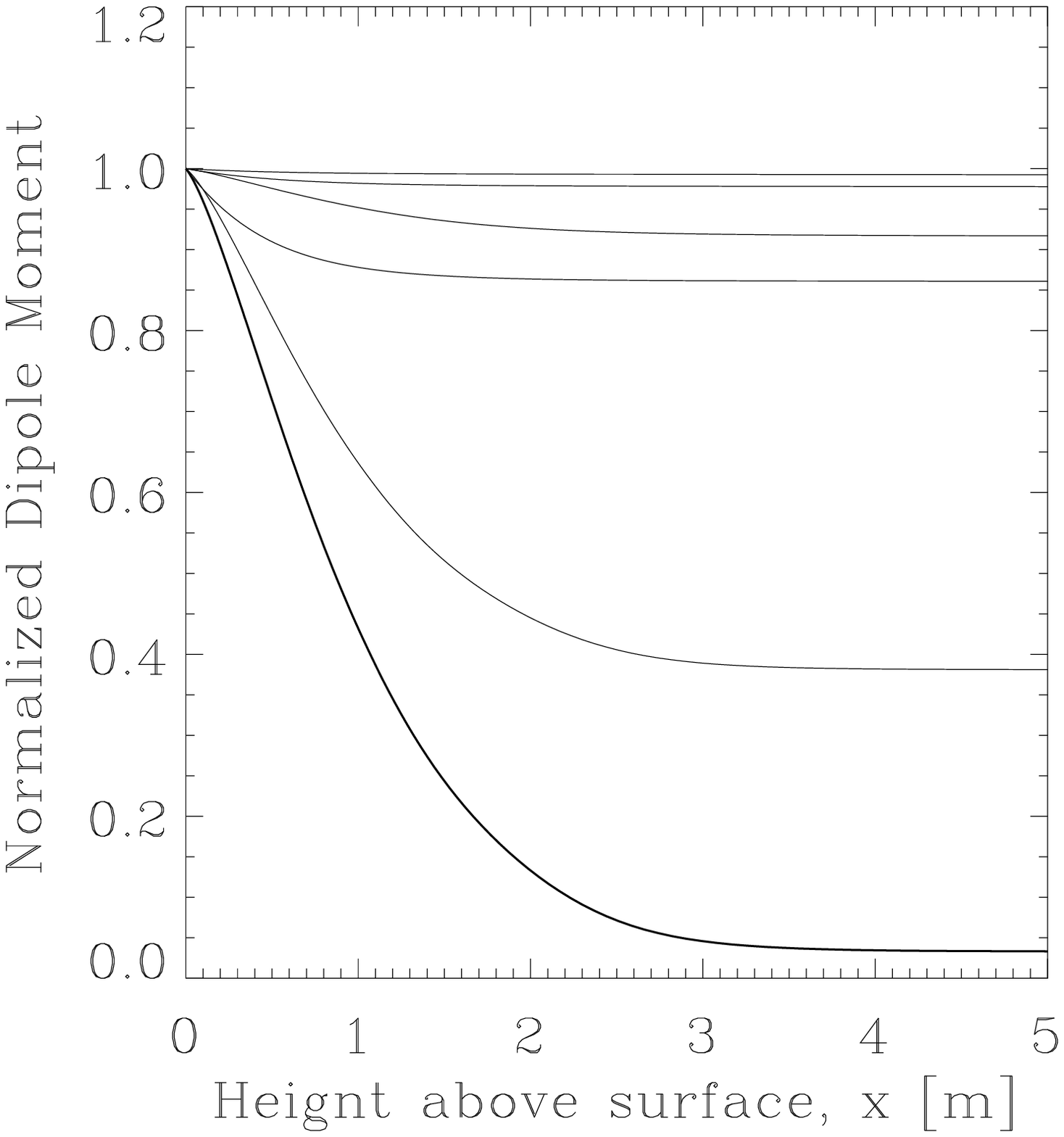} 
 \end{tabular}
 \begin{tabular}{c}
  (b) \\
  \includegraphics[height=60mm]{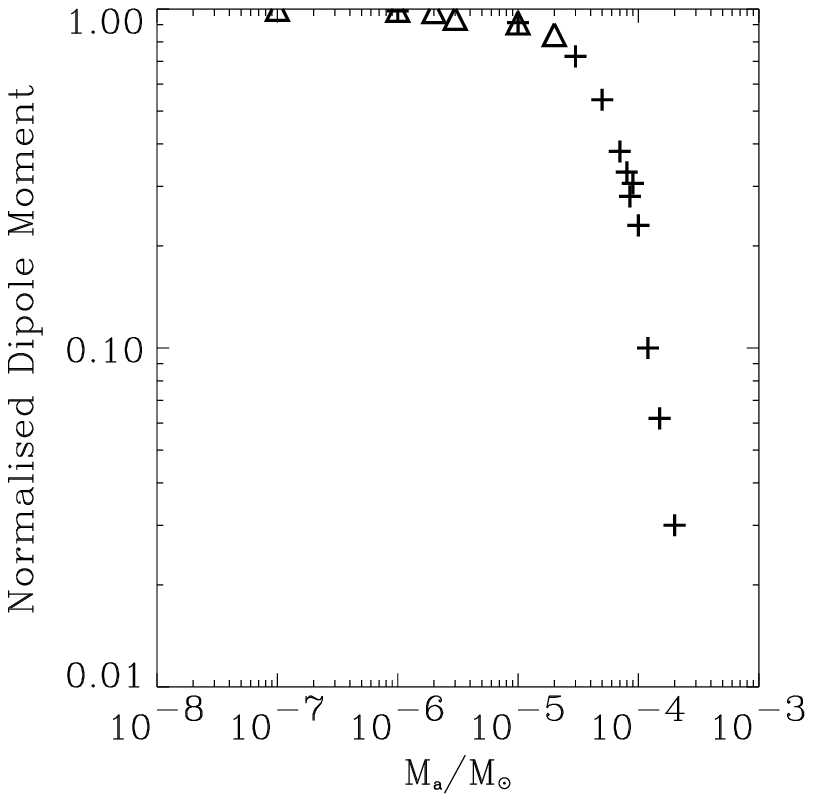} 
 \end{tabular}
 \begin{tabular}{c}
  (c) \\
  \includegraphics[height=60mm]{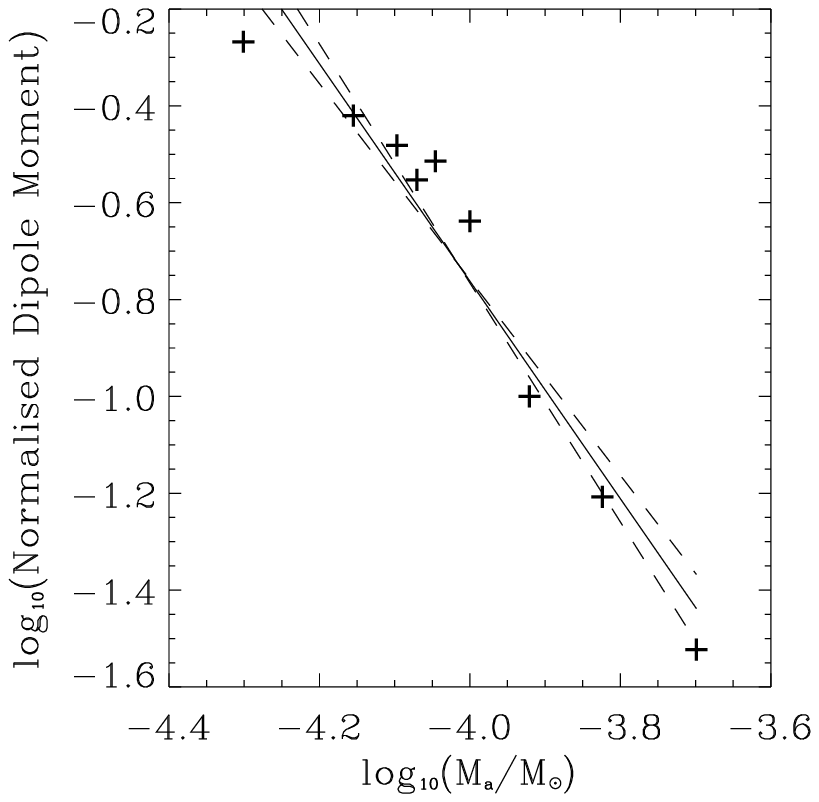} 
 \end{tabular}
\caption{\small
(a) Dipole moment $|\vv{m}|$ as a function of altitude
$x = r-R_{*}$, normalised to $|\vv{m}|$ at $r = R_{*}$,
for
$M_{\rm a}/\Msun = 
10^{-6}\,$(\emph{top}),\, $3\times 10^{-6},\, 10^{-5},\, 3\times 10^{-5},\, 10^{-4},\, 3\times 10^{-4}$ (\emph{bottom}).
(b) Dipole moment $|\vv{m}|$ as a function of
accreted mass $M_{\rm a}$,
normalized to the dipole moment $|\vv{m}_{\rm i}|$ before accretion,
for
$b = 3$ (\emph{crosses}) and
$b = 10$ (\emph{triangles}).
(c) Power law fit (\emph{solid}), with $1\sigma$ errors
(\emph{dashed}), to dipole moment as a function of
$M_{\rm a}$ for
$M_{\rm a} > 2\times 10^{-5}\Msun$.
}
\label{fig:dipole}
\end{figure}
\end{center}
\subsection{Onset of spreading}
\label{sec:spreading}

\citet{bro98} showed that, for an initially vertical field
and neglecting the stress from compressed equatorial flux,
the condition for spreading
is given by
$\alpha = B^{2}/2\mu_{0}p \leq x_{0}/R_{\rm cap} \approx 0.01$,
where
$R_{\rm cap} = (R_{*}^{3}/R_{\rm a})^{1/2}$ is the polar cap
radius.
Given that
$\alpha_{\rm min} = B_{*}^{2}/(2\mu_{0}p_{\rm max}) \approx
0.27(B_{*}/10^{8}\,{\rm T})^{2}(M_{\rm a}/10^{-12}\Msun)^{-1}$,
it follows that the accreted matter distorts the magnetic
field negligibly for 
$M_{\rm a} \lesssim 3.7 \times 10^{-9}\Msun$
(i.e.\ $\alpha_{\rm min} \gtrsim 0.01$).
Above this value, progressively more distortion takes place
for
$M_{\rm a} = 10^{-9}, 10^{-7}, 10^{-5}\Msun$, as shown
in Figure \ref{fig:varyMa}.
The curved field lines have a large tangential component, previously negligible,
which increases the magnetic field strength |$\vv{B}$| substantially for
$M_{\rm a} \gtrsim M_{\rm c}/a = 8\times 10^{-11}\Msun$,
as predicted by the 
the Green function analysis in 
Appendix \ref{greenappendix3}.
This induced magnetic pressure balances the overpressure of the accreted material.
However, the effect on the magnetic dipole moment is negligible
($< 1\%$) due to countervailing magnetic stresses
from the compressed field at the equator;
the magnetic radius of curvature
is less than $x_0$ until
$M_{\rm a}$ exceeds $\approx 1.4\times 10^{-6}\Msun$,
as proved in Appendix \ref{greenappendix3}.
Note that the compression of field lines as $|\vv{B}|$
increases is imperceptible in Figure \ref{fig:varyMa}
because of the extent of
the horizontal axis;
$|\vv{B}|$ 
increases predominantly due to the $B_{\theta}$ component.
The top of the boundary layer is roughly where
$B_{\theta}$ vanishes, i.e.\ at an altitude
$x = x_{0} \ln[(a+1)/(1 - M_{\rm c}/M_{\rm a})] \approx  5.3$ m.
for $M_{\rm a} < M_{\rm c}$.

We zoom in on the pole to compare with \citet{lit01}
in Figure \ref{fig:comparelitwin}.
Our results differ because
\citet{lit01} has a free boundary at the polar cap edge and ignores
the $\theta$ terms in the Grad-Shafranov equation, while we impose
north-south symmetry at the equator, with no condition at the polar cap
edge.  This allows the compressed magnetic field equatorward
of the polar cap to push back on the polar flux tube.
Furthermore, we prescribe $dM/d\psi$ and calculate $F(\psi)$
instead of prescribing $F(\psi)$.
We observe curvature comparable to \citet{lit01} 
for $M_{\rm a} \approx 10^{-8}\Msun$ and thus the ballooning instability
may be relevant, but a detailed calculation is beyond the scope of this paper.

An order of magnitude estimate of $M_{\rm c}$, the mass required to
buckle the magnetic field, can be obtained in the following way.
The hydrostatic pressure at the base of the accreted column is given by
$p_{\rm max} = c_{\rm s}^2\rho_{\rm max} = c_{\rm s}^2 M_{\rm a}b^2/(2\pi R_{*}^{2}x_0)$,
i.e.\ the weight per unit area of the mass $M_{\rm a}$ spread
over approximately two hemispheres.
This pressure is balanced by the tension of the magnetic field
compressed into a layer of width
$L\approx 6$ m along the surface and at the equator
(see Figure \ref{fig:polar}).
In the layer, we have
$B_{1} \approx B_{*}R_{*}/L \approx 10^{12}$ T\, by flux
conservation.
Hence the hydrostatic and magnetic pressures balance for
$M_{\rm a} = 2 B_{*}^2 R_{*}^4 x_{0}/(c_{\rm s}^2 \mu_0 L^2)
\approx 2\times 10^{-6}\Msun$
in accord with the numerical results.

\subsection{Reduction of the magnetic dipole moment}
\label{sec:dipolereduce}
The magnetic dipole moment 
\begin{equation}
\label{eqn:dipole}
 |\vv{m}| =
 \frac{3r^3}{4}
 \int^1_{-1} d(\cos\theta)\, \cos\theta
 B_r(r,\theta)
\end{equation}
is plotted as a function of
$r$ in Figure \ref{fig:dipole}(a).
The screening currents are confined to a thin layer
above the stellar surface; 
$|\vv{m}|$ is essentially constant with $r$ above this layer.
The layer is compressed as $M_{\rm a}$ increases,
with half-width comparable to $x_0$ for
$M_{\rm a} \approx 10^{-4}\Msun$.
The asymptotic value of $|\vv{m}|$
also decreases as $M_{\rm a}$ increases,
as expected;
equatorward hydromagnetic spreading drags magnetic
flux away from the pole,
and $|\vv{m}|$ is sensitive to $B_{r}$ near the pole
through (\ref{eqn:dipole}).

Figure \ref{fig:dipole}(b) is a plot of $|\vv{m}|$
versus $M_{\rm a}$.
For $M_{\rm a} \lesssim M_{\rm c} = 1.2\times 10^{-6}\Msun$,
$|\vv{m}|$ decreases proportional to
$(1 - M_{\rm a}/M_{\rm c})$, as predicted analytically
in Section \ref{smallma}.
For $M_{\rm a} \gtrsim M_{\rm c}$,
we obtain the empirical relation
\begin{equation}
\label{eqn:empiricaldipole}
|\vv{m}|/|\vv{m}_{\rm i}| = (M_{\rm a}/4.6\times 10^{-5}\Msun)^{-2.25\pm0.22}
\end{equation}
by fitting a power law to the numerical results for
$M_{\rm a} \geq 5\times 10^{-5}\Msun$,
as in Figure \ref{fig:dipole}(c).
However, 
(\ref{eqn:empiricaldipole}) cannot be extrapolated reliably
to the regime
$M_{\rm a} \gtrsim 10^{-4}\Msun$ for two reasons.
First,
our numerical scheme is limited by the steepness of
gradients in the source term of
(\ref{mhdstaticF}).  Physically, these scale as
the hydrostatic pressure, with
$dF/d\psi \propto M_{\rm a}b$ as shown
in Appendix \ref{greenappendix}.
We encounter 
convergence errors above $50\%$
for $M_{\rm a}b \gtrsim 3\times 10^{-4}\Msun$.
Second,
it is shown in Section \ref{sec:bubbles}
that magnetic bubbles, disconnected from the stellar
surface, are created for
$M_{\rm a}b \gtrsim 10^{-4}\Msun$,
leading to
the steep dependence of
$|\vv{m}|$ on $M_{\rm a}$ in (\ref{eqn:empiricaldipole}).
A word of caution:
when bubbles appear,
it is unclear how to interpret
$|\vv{m}| = |\vv{m}|_{\rm star} + |\vv{m}|_{\rm bubble}$.
In reality, one has $|\vv{m}|_{\rm bubble} = 0$ as
$r\rightarrow\infty$, because the flux surfaces of the bubble
are closed.
However, at $r = R_{\rm m}$, the ingoing and outgoing
flux tubes of the bubble do not cancel perfectly
and one finds  $|\vv{m}|_{\rm bubble}\neq 0$;
bubble-related currents
outside the solution domain ($r > R_{\rm m}$),
need to be included
in order to recover $|\vv{m}|_{\rm bubble} = 0$.

\begin{center}
\begin{figure*}
\begin{tabular}{cc}
\includegraphics[height=65mm]{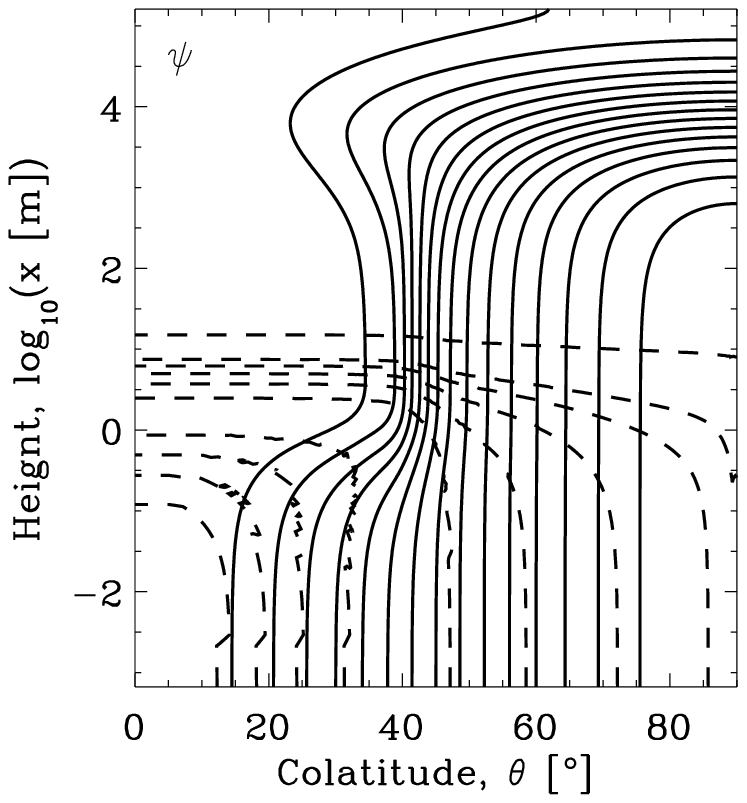}
&
\includegraphics[height=65mm]{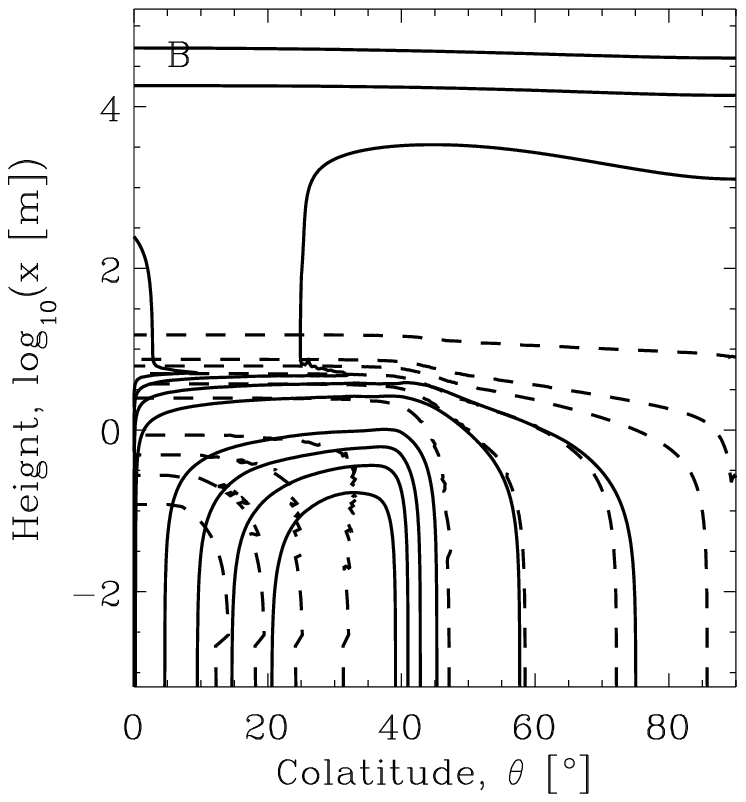}
\\
\includegraphics[height=65mm]{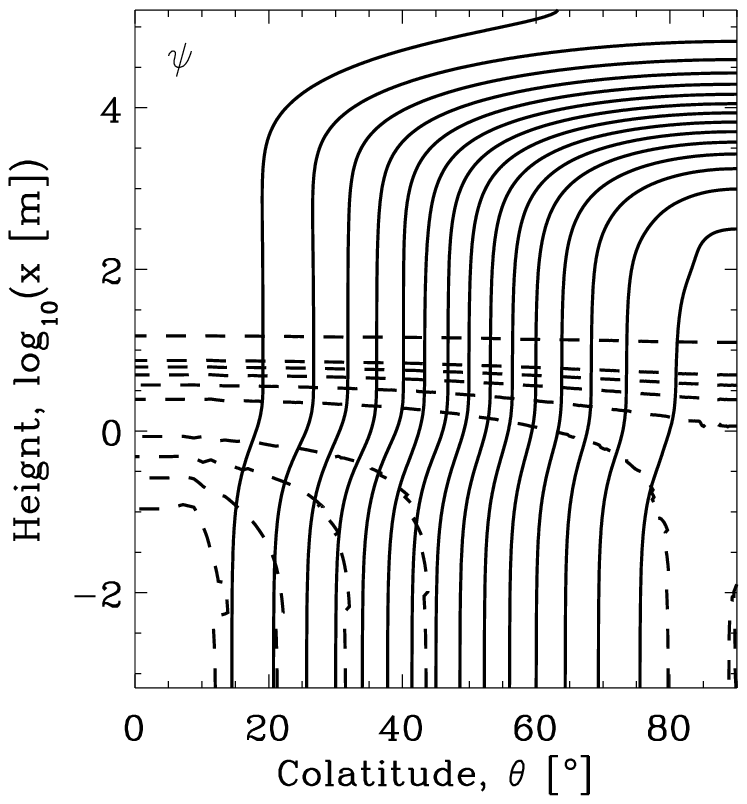}
&
\includegraphics[height=65mm]{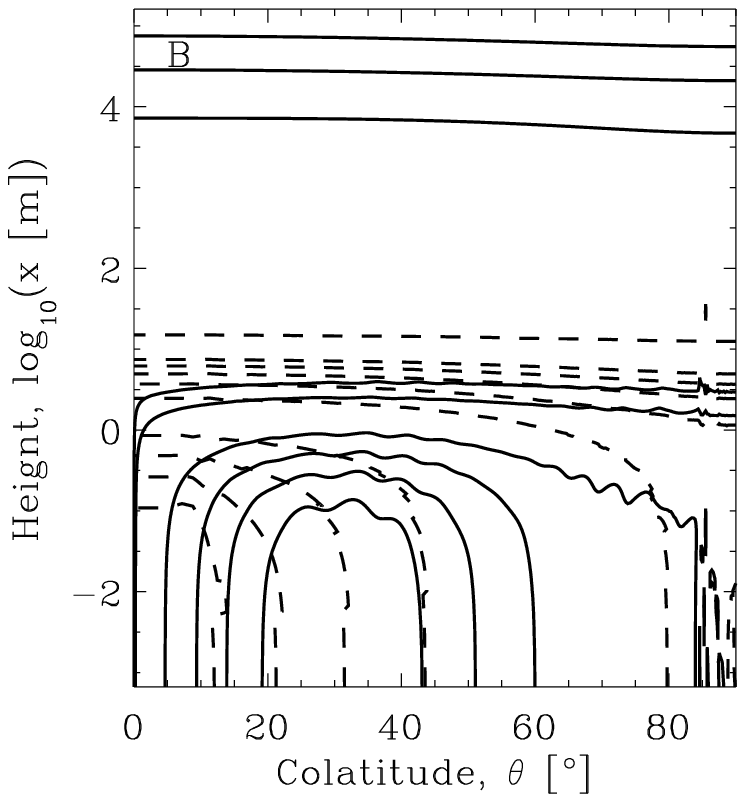}
\end{tabular}
\caption{\small
Hydromagnetic equilibria for $M_{\rm a} = 10^{-5}\Msun$,
with
$b = 10$ (\emph{top}) and
$b = 3$ (\emph{bottom}).
Contours of 
constant $\psi$ (\emph{left}) and
$|\vv{B}|$  (\emph{right}) are displayed, with
$|\vv{B}|_{\rm max} = 4.06 \times 10^{11} {\rm T}$ ($b = 10$) and
$1.65\times 10^{11} {\rm T}$ ($b = 3$).
We find
$|\vv{m}| \approx 0.9|\vv{m}_{\rm i}|$ in both cases.
}
\label{fig:varyb}
\end{figure*}
\end{center}

\subsection{Polar cap radius}
\label{sec:polarcap}

We now discuss the effect on $|\vv{m}|$ of varying
$b = \psi_{\rm a}/\psi_{*}$,
or equivalently the
polar cap radius
$R_{\rm cap} = R_{*}\sin^{-1}(b^{-1/2})$.
Although $R_{\rm cap}$ is not known exactly without
a detailed
model of the flow of matter from the accretion disc to the
stellar surface \citep{aro84},
estimates of its size from (\ref{eq:alfven})
are typically of order $1\, {\rm km}$
\citep{lit01},
i.e. $b \geq 100$, for $B_{*} = 10^{8}$ T
\citep{aro84}.
Figure \ref{fig:varyb}
illustrates the magnetic configuration obtained for 
$b = 3$ 
and $b = 10$, 
and
$|\vv{m}|/|\vv{m}_{\rm i}|$ is plotted versus
$M_{\rm a}$ in Figure \ref{fig:dipole}(b),
denoting $b = 3$ by crosses and $b = 10$
by triangles.
Note how the equilibrium state changes with $b$.
The mass-flux distribution
$d{M}/d{\psi} \propto \exp(-\psi/\psi_{\rm a})$
implies a surface pressure distribution
${F}({\psi}) \propto b\exp(-{\psi/\psi_{\rm a}})$,
so 
larger $b$ means steeper pressure gradients.
(Numerical difficulties set in for $b = 30$,
which can be partly
alleviated by stretching the coordinates around
$\psi_{\rm a}$.)
Importantly, however, we find
$M_{\rm c}$ and the dipole
moment $|\vv{m}|$ are independent of $b$.
This is supported by the Green function analysis in 
Appendix \ref{greenappendix}, except when
$dM/d\psi$ depends
explicitly on $b$,
e.g.\
$d{M}/d{\psi} \propto b(1-{\psi}/b\psi_{\rm a})^{1/2}$
implies $|\vv{m}| \propto b^{-1}$.

\begin{center}
\begin{figure}
\centering
\includegraphics[height=65mm]{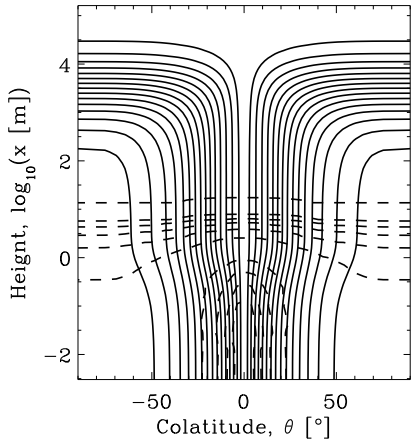} 
\caption{\small
Equilibrium magnetic configuration in the northern hemisphere,
showing contours of constant $\psi$ ({\em solid})
and $\rho$ ({\em dashed}).
Equations (\ref{mhdstaticF}) and (\ref{fpsi})
are solved in the domain
$|\theta|\leq\pi/2$ here,
compared to $0\leq\theta\leq\pi/2$ in earlier figures,
in order to test the validity of the
$\psi = 0$ boundary condition at
$\theta = 0$.
}
\label{fig:equator}
\end{figure}
\end{center}

A related issue is whether $|\vv{m}|$ is affected in
an unrealistic way by the boundary condition $\psi = 0$
at $\theta = 0$ in the examples presented so far.
It is conceivable, for example,
that the $\psi = 0$
line is unstable (`on a knife edge'),
while neighbouring field lines are peeled away by accretion,
unless it is forced to remain rectilinear artificially.
As it happens, however, this is not the case.
Figure \ref{fig:equator} shows the output of an experiment where the grid
extends from equator to equator 
($|\theta| < \pi/2$), and
no boundary condition is imposed at
$\theta = 0$.
Clearly,
the magnetic field
and density profiles remain symmetric about the magnetic pole,
with $\psi = 0$ at $\theta = 0$ emerging naturally,
while $|\vv{m}|$ is essentially unchanged.

\subsection{Buoyant magnetic bubbles}
\label{sec:bubbles}

\begin{center}
\begin{figure}
\includegraphics[height=65mm]{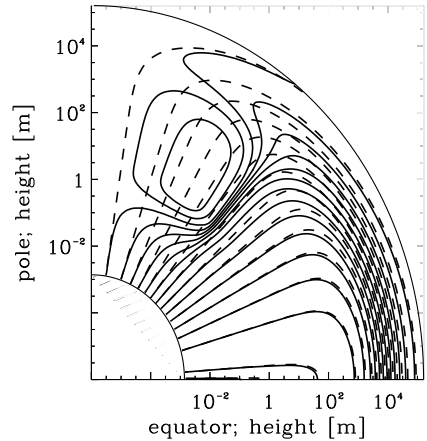}
\includegraphics[height=65mm]{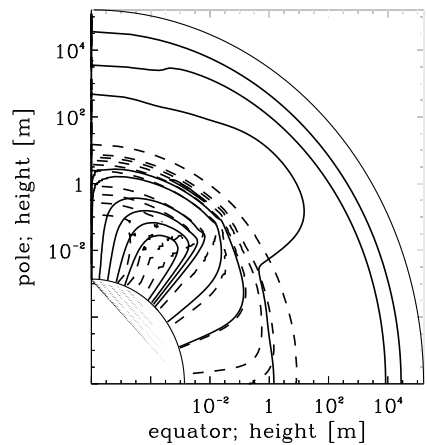} 
\caption{\small
Magnetic configuration
for $M_{\rm a} = 2\times 10^{-5}\Msun$, $b = 10$,
showing the creation of a bubble.
We plot contours of constant $\psi$ (\emph{top})
for initial (\emph{dashed})
and final (\emph{solid}) states, and
final $|\vv{B}|$ contours
$\eta |\vv{B}|_{\rm max}$ (\emph{bottom, solid})
and $\rho$ contours (\emph{dashed})
$\eta\rho_{\rm max}$;
$\rho_{\rm max} = 5\times 10^{17}\, {\rm kg m}^{-3}$,
$B_{\rm max} = 6.3\times10^{11}$\, T, and
$\eta = 0.8,\, 0.6,\, 0.4,\, 0.2,\, 0.01,\, 0.001,\, 10^{-4},\,
10^{-5},\, 10^{-6},\, 10^{-12}$.
}
\label{fig:M5b}
\end{figure}
\end{center}

From the sequence of panels in Figure \ref{fig:varyMa}
($M_{\rm a}/\Msun = 10^{-9},\, 10^{-7},\, 10^{-5}$),
we observe that the magnetic field becomes increasingly
distorted as $M_{\rm a}$ increases.
Eventually,
for $M_{\rm a} \gtrsim 10^{-5}\Msun$,
closed magnetic bubbles
are created that are disconnected topologically 
from the surface of the star.
This phenomenon is illustrated in
Figure \ref{fig:M5b}.
At the value of $M_{\rm a}$ where a bubble is first created,
a magnetic neutral point (Y point) is observed to form
on a field line near, but not at the pole.
The bubble closes at $r < R_{\rm m}$ in our simulation,
but this may be a result of the approximate
free boundary condition
$\partial\psi/\partial\theta(R_{\rm m},\theta) = 0$;
in reality, it may connect to the accretion disc.

Bubbles correspond to a loss of equilibrium,
analogous 
to that which occurs during 
eruptive solar phenomena \citep{kli89},
where no simply connected hydromagnetic
equilibrium exists.
In the Grad-Shafranov boundary problem, the source term
$\propto F^{\prime}(\psi)$ in (\ref{mhdstaticF})
increases with $M_{\rm a}b$, boosting
$\Delta^2\psi$ and hence $\psi$ above the surface
(to balance the weight of the added material through the
Lorentz force).
Above a critical value of $M_{\rm a}$,
flux surfaces are created with
$\psi < 0$ or $\psi > \psi_{*}$, which are
disconnected from the star and
either form closed loops or are anchored
`at infinity' (here the accretion disc).
This is shown explicitly by (\ref{eq:greenpsi})
in the special case
$F^{\prime}(\psi) =$ constant.
From our numerical results, we conclude that the
critical $M_{\rm a}$ for bubble creation satisfies
\begin{equation}
M_{\rm a} \geq 10^{-4}b^{-1}\Msun\, .
\label{eqn:bubblebreak}
\end{equation}
Note that bubble creation is a topological imperative.
It is not the result of
a hydromagnetic instability
e.g.\ interchange or
Rayleigh-Taylor
\citep{ber58,par66,mou74}.

Are the bubbles merely numerical artefacts
\citep{bro98}?
No.
Equation (\ref{eq:greenpsi})
demonstrates explicitly that flux surfaces with
$\psi < 0$ or $\psi > \psi_{*}$ are created for
$M_{\rm a}$ satisfying (\ref{eqn:bubblebreak}),
at least for
$F^{\prime}(\psi) =$ constant.
A more subtle
issue is whether bubbles
are the by-product of an artificial
assumption in our idealized calculation.
For example, if submergence of accreted material
were permitted, it might
reduce the pressure gradients that produce
the bubbles;
on the other hand, ohmic dissipation would facilitate
detachment of bubbles in a pinched, Y-point
configuration.

On some runs, bubbles
appear and disappear
during the iteration process.
This happens because the mass-flux distribution
is not conserved inside a bubble,
although the code attempts to maintain flux freezing
at the edge.
If the route to convergence is a rough 
proxy for time-dependent behaviour,
as argued by \citet{mou74} for iterative relaxation
algorithms,
the appearance and disappearance of bubbles may
represent evidence --- though not proof ---
of transient evolution in reality.

As the bubbles are disconnected topologically from
flux surfaces anchored to the star and
accretion disc, they do not contain any accreted
material (in the ideal-MHD limit of zero cross-field
transport) and are lighter than their surroundings.
It is therefore possible that they rise buoyantly
and ultimately escape the magnetosphere
of the neutron star.
This possibility cannot be investigated rigorously in
the context of the equilibrium
calculations in this paper;
it is considered qualitatively in Section
\ref{sec:buoyantbubbles}.

\section{TIME-DEPENDENT EFFECTS}
\label{sec:discussion}

In this section, we discuss critically
(but qualitatively)
how the results of this paper may be affected by time-dependent
processes that cannot be modelled 
by a quasi-static sequence of hydromagnetic equilibria.
We consider Parker instabilities in Section 5.1, ohmic dissipation
in Section 5.2, and the buoyant rise of magnetic bubbles
in Section 5.3.

\subsection{Hydromagnetic instabilities}
The computed equilibria are manifestly distorted.
Buoyancy of the
compressed magnetic flux can drive long-wave, slow MHD modes
that overturn the accreted matter on the Alfv\'en time-scale $\tau_{\rm A}$,
as in the global Rayleigh-Taylor
instability of the Galactic magnetic field
\citep{par66,mou74}.
When the accreted matter bends the polar magnetic
field, there exists a significant component of magnetic field
perpendicular to gravity,
a condition for the onset of the Parker instability.
In a plane-parallel geometry,
wavelengths longer than
$\Lambda = 4\pi x_0/(2\alpha+1)$ are unstable \citep{mou74},
where $\alpha = {B}^2/(2\mu_0 p)$.
The geometry of
an accreting neutron star
is far from plane-parallel.
Nevertheless, in hydromagnetic equilibrium, one has
$\alpha \sim 1$ locally in the boundary layer, yielding
$\Lambda = 2$\, m.
The failure to converge at large $M_{\rm a}$ is also
a hint, though not a proof, that the Parker instability
may operate.
\citet{mou74} advanced a similar convergence-based
argument for the stability of a
stratified vertical column
with periodic boundaries.

\subsection{Ohmic dissipation}
\label{ohmicloss}

Our calculations are performed under
the assumption
of infinite conductivity and hence 
flux freezing.
In reality, the accreted matter
is resistive due to electron-phonon
and electron-impurity scattering
\citep{bro98,cum01},
potentially enhanced by accretion-induced heating
\citep{rom90,urp95}.

The ohmic dissipation time-scale for a flux tube of
width $L$ is given by
$\tau_{\rm d} = \mu_{0}\sigma L^2$,
where $\sigma$ is the electrical conductivity.
For typical conditions, we take
$\sigma \approx 10^{22} {\rm s}^{-1}$
and hence obtain
$\tau_{\rm d} \approx 10^{14} (L/R_\ast)^2\,{\rm s}$.
In comparison, the flow time-scale is given by
$\tau_{\rm f} = 4\pi R_\ast^2 \rho L /
\dot{M}_{\rm a},
$
where $\dot{M}_{\rm a}$ is the accretion rate.
For
$\rho = 4\times 10^{14}\, {\rm kg\, m}^{-3}$, $L = 6$\, m
(Section \ref{sec:spreading}), and
$\dot{M}_{\rm a} = 1\times 10^{-8}\Msun$\, yr$^{-1}$,
we find
$\tau_{\rm f} = 1\times 10^{2}$\, yr.
Therefore, for the length-scales characteristic
of the compressed flux layer, we have
$\tau_{\rm f} > \tau_{\rm d}$ and
magnetic flux diffuses through the accreted material,
broadening the compressed flux layer until it is
thick enough ($L \approx 600$\, m) that
$\tau_d \approx \tau_f$ and further thickening ceases.
\citet{bro98} showed that
$\tau_{\rm f}/\tau_{\rm d}$
depends only on $\dot{M}_{\rm a}$
and not on depth in the crust.

Note that the buried flux is resurrected
on the time-scale $\tau_{\rm d}$, after accretion stops.
As the noted dipole field reasserts itself,
$|\vv{m}|$ increases.
A full analysis of this process is left to future work.

\subsection{Buoyant bubbles}
\label{sec:buoyantbubbles}

Closed magnetic bubbles have $\rho = 0$ within
and are lighter than their surroundings, as discussed in
Section \ref{sec:bubbles}.
They tend to rise buoyantly at the local Alfv\'en speed
$v_{\rm A} = (B^2/\mu_{0}\rho)^{1/2} = c_{\rm s}(2\alpha)^{1/2}$,
and hence escape the magnetosphere in $\approx 2$\, yr.
Assuming an accretion rate of $10^{-8}\Msun\, {\rm yr}^{-1}$,
it takes
$\approx 10^{4}b^{-1}(\dot{M}_{\rm a}/10^{-8}\Msun\, {\rm yr}^{-1})^{-1}$\, yr
to accrete enough mass
to create one bubble.
Moreover, a typical bubble encloses
$\approx \psi_{\rm a}$ of magnetic flux.
Hence we conclude that magnetic flux is being expelled episodically from the magnetosphere at an average rate of
$10^{11}(\psi_{*}/10^{16}\, {\rm T m}^{2}) \
(\dot{M}_{\rm a}/10^{-8}\Msun\, {\rm yr}^{-1})^{-1}$ T\, m$^{2}$\, yr$^{-1}$.
Note that expelled flux is subsequently replenished by the current
deep in the star
(since $\psi$ is fixed at the surface).

\section{CONCLUSION}

Observations of low-field binary neutron stars and recycled
pulsars imply that the magnetic dipole moment of a neutron
star is reduced by accretion.
In this paper, we undertake a self-consistent analysis
(numerical and analytic) of one mechanism that may account
for the reduction observed:
polar magnetic burial in the ideal-magnetohydrodynamic regime.
Our analysis has several new features.
(i) Flux freezing is strictly enforced when connecting the
final and initial magnetic configurations
by solving
self-consistently for the mass-flux distribution rather
than specifying it ad hoc
(see Figure \ref{fig:compareF}).
(ii) The Lorentz force due to equatorial magnetic field
lines compressed by equatorward hydromagnetic spreading is
included when calculating the confinement
of the polar accreted column.
(iii) Numerical methods are developed for treating accreted masses
up to $\approx 10^{-4}\Msun$,
(cf. $10^{-10}\Msun$ in previous work),
where the field is dramatically distorted and high-order
multipoles dominate.

We report two key results.
(i) $M_{\rm a} \gtrsim 10^{-5}\Msun$ must be accreted
in order to reduce significantly the magnetic
dipole moment $|\vv{m}|$ of the star,
contrary to previous estimates
($M_{\rm a} \approx 10^{-10}\Msun$)
which neglected equatorial magnetic stresses.
For small $M_{\rm a}$, we find
$|\vv{m}| = |\vv{m}_{\rm i}|(1 - M_{\rm a}/M_{\rm c})$,
with
$M_{\rm c} = 1.2\times 10^{-6}\Msun$,
(cf.\ \citealt{shi89}).
For $M_{\rm c} \lesssim M_{\rm a} \lesssim 10^{-4}\Msun$,
we find
$|\vv{m}| = |\vv{m}_{\rm i}|(M_{\rm a}/4.6\times 10^{-5}\Msun)^{2.25\pm 0.22}$.
(ii) When enough mass is accreted,
such that $M_{\rm a} \geq 10^{-4}b^{-1}\Msun$,
the hydrostatic pressure gradient generates flux surfaces with
$\psi < 0$ or $\psi > \psi_{*}$,
creating closed magnetic bubbles that are
disconnected topologically from the stellar surface.
The bubbles are valid solutions of the Grad-Shafranov boundary problem,
as confirmed by analytic,
Green function calculations;
they are not numerical artefacts
or fingerprints of hydromagnetic instabilities
(e.g.\ Parker).

Several of our assumptions need to be relaxed in future work,
including
(i) perfect conductivity,
(ii) an impenetrable stellar surface,
and
(iii) axisymmetry
(which tends to suppress hydromagnetic instabilities, for example).
Finally, the uniqueness of the hydromagnetic equilibria we compute
numerically is yet to be established.

Our equilibria serve as useful starting points for
exploring the stability of the magnetic configuration
during and after accretion.
Our theoretical results will be
tested against observational data from binary
neutron stars and recycled pulsars in a companion paper.

\bibliographystyle{mn2e}
\bibliography{b}

\appendix

\section{Analytic solution of the Grad-Shafranov problem}
\label{greenappendix}

In this appendix, we solve the Grad-Shafranov equation
(\ref{mhdstaticF}), together with the boundary conditions
(\ref{boundary}),
by a Green function approach.
\subsection{Green theorem for the Grad-Shafranov operator}
\label{greenappendix1}
An operator
$L\psi = \nabla^2\psi + \vv{b}\cdot\nabla\psi + c\psi$,
acting on a function $\psi$, possesses an adjoint
$L^*\psi = \nabla^2\psi - \nabla\cdot(\vv{b}\psi) + c\psi$.
The Grad-Shafranov operator $L = \mu_{0}r^2\sin^2\theta\Delta^2$,
defined by (\ref{gs}) in
spherical polar coordinates,
has 
$\vv{b} = -2r^{-1}(\hat{\vv{e}}_{r} + \cot\theta\hat{\vv{e}}_{\theta})$
and is not self-adjoint (cf. $\nabla^2$).
Letting $G$ and $G^{*}$ be the Green functions associated with the
operators $L$ and $L^{*}$ respectively,
\begin{equation}
LG(\vv{x},\vv{x}^\prime) = \delta(\vv{x} - \vv{x}^{\prime})\, ,
\label{Gdelta}
\end{equation}
\begin{equation}
L^{*}G^{*}(\vv{x},\vv{x}^\prime) = \delta(\vv{x} - \vv{x}^{\prime})\, ,
\label{Gsdelta}
\end{equation}
related by the reciprocity relation
$G^{*}(\vv{x},\vv{x}^\prime) = G(\vv{x}^{\prime},\vv{x})$,
we arrive at the Lagrange identity
\begin{equation}
\psi L^{*}G^{*} - G^{*}L\psi = \nabla\cdot(\psi\nabla G^{*} - G^{*}\nabla\psi + \vv{b}\psi G^{*}).
\label{adjoint}
\end{equation}
Upon integrating (\ref{adjoint}) over a volume $V$, bounded by a surface $S$, and using
the divergence theorem, we obtain
\begin{equation}
\begin{split}
 \int_V (\psi L^{*}G^{*} - G^{*}L\psi) dV
\quad\quad\quad\quad\quad\quad\quad\quad\quad \\
  =
 \int_S (\psi\nabla G^{*} - G^{*}\nabla\psi + \vv{b}\psi G^{*})\cdot\hat{\vv{n}}dS,
\label{greentheorem}
\end{split}
\end{equation}
where $\hat{\vv{n}}$ is the unit vector normal to $S$.
Given a boundary value problem
$L\psi(\vv{x}) = Q(\vv{x})$
in the volume $V$, with $\psi$ given on the boundary $S$,
we combine (\ref{Gsdelta}) and (\ref{greentheorem}) to obtain
\begin{equation}
\begin{split}
\psi(\vv{x})
 &= \int_V d^3\vv{x}^{\prime} G^{*} Q \\
 & \quad + \int_S d^2\vv{x}^{\prime}\hat{\vv{n}}\cdot(\psi\nabla G^{*}-G^{*}\nabla\psi+\vv{b}\psi G^{*}).
\end{split}
\end{equation}

\subsection{Green function for the Grad-Shafranov equation}
\label{greenappendix2}
We wish to solve (\ref{mhdstaticF}) for $\psi$ in $r\geq R_{*}$ subject to
Dirichlet boundary conditions (\ref{boundary}) on $\psi$ at $r = R_{*}$
and $r\rightarrow\infty$.
In cylindrical symmetry, the
volume and surface integrals in (\ref{psisolution}) reduce to surface and line
integrals respectively.
It is convenient to make the substitution $\mu = \cos\theta$, whereupon
(\ref{mhdstaticF}) becomes
\begin{equation}
\begin{split}
\frac{\partial^2\psi}{\partial r^2} + \
\frac{(1-\mu^2)}{r^2}\frac{\partial^2\psi}{\partial\mu^2} \
 = \mu_0 r^2(1-\mu^2)\frac{dF(\psi)}{d\psi} \\
 \times \exp(-{\phi_{0}}/{c_{\rm s}^2}-GMr/R_{*}^2c_{\rm s}^2)
\label{gsmu}
\end{split}
\end{equation}
and $dF/d\psi$ is a function of $r$ and $\mu$ through $\psi(r,\mu)$.

We redefine $L$ to be the operator on the left-hand side of (\ref{gsmu}), and
$Q(r,\mu)$ to be the source term on the right-hand side,
known explicitly once $F(\psi)$ is known.
The Green function $G$ for $L$ satisfies
\begin{equation}
\frac{\partial^2G}{\partial r^2} + \\
\frac{(1-\mu^2)}{r^2}\frac{\partial^2G}{\partial\mu^2} \\
= \frac{1}{r^2}\delta(r - r^{\prime})\delta(\mu - \mu^{\prime})
\end{equation}
and the Green function $G^{*}$ for $L^{*}$ satisfies
\begin{equation}
\begin{split}
\frac{\partial^2G^{*}}{\partial r^2} + \
\frac{(1-\mu^2)}{r^2}\frac{\partial^2G^{*}}{\partial\mu^2} \
+ \frac{4}{r}\frac{\partial G^{*}}{\partial r} \
- \frac{4\mu}{r^2}\frac{\partial G^{*}}{\partial\mu} \\
= \frac{1}{r^2}\delta(r - r^{\prime})\delta(\mu - \mu^{\prime}).
\label{eqn:gsmuGrepeat}
\end{split}
\end{equation}
Equation (\ref{eqn:gsmuGrepeat}) is separable.
We expand the solution 
in terms of orthogonal Gegenbauer polynomials $C^{3/2}_{\ell}(\mu)$, viz.
\begin{equation}
\begin{split}
G(r,\mu, r^{\prime},\mu^{\prime})
 &=
\sum_{\ell = 0}^{\infty} g_{\ell}(r, r^{\prime}) \\
 & \quad \times (1-\mu^2)C_{\ell}^{3/2}(\mu^{\prime})C_{\ell}^{3/2}(\mu),
\end{split}
\end{equation}
with
\begin{equation}
\label{eq:smallg}
\frac{d^2g_{\ell}(r,r^{\prime})}{dr^2} -\frac{\ell(\ell+1)}{r^2}g_{\ell}(r,r^{\prime}) = r^{-2}\delta(r - r^{\prime}).
\end{equation}
The Gegenbauer polynomials satisfy
\begin{equation}
\begin{split}
(1-\mu^2)\frac{d^2}{d\mu^2}\left[(1-\mu^2)C^{3/2}_{\ell-1}(\mu)\right] & \\
 + \,\,   \ell(\ell+1)[(1-\mu^2)C^{3/2}_{\ell-1}(\mu)] =& \,\, 0
\end{split}
\end{equation}
and are related to associated Legendre polynomials via
$P_{\ell}^{1}(\mu) = -(1-\mu^2)^{1/2} C_{\ell-1}^{3/2}(\mu)$.
The first few are listed for reference:
$C_{0}^{3/2}(\mu) = 1$,
$C_{1}^{3/2}(\mu) = 3\mu$,
$C_{2}^{3/2}(\mu) = \frac{3}{2}(5\mu^2 -1)$,
$C_{3}^{3/2}(\mu) = \frac{5}{2}(7\mu^3 - 3\mu)$,
$C_{4}^{3/2}(\mu) = \frac{15}{4}(21\mu^4 - 14\mu^2 +1)$,
$C_{5}^{3/2}(\mu) = \frac{3}{8}(231\mu^5 - 210\mu^3 + 35\mu)$.
They satisfy an orthogonality condition:
\begin{equation}
\int_{-1}^{1}(1-\mu^2)C_{\ell}^{3/2}(\mu)C_{\ell^{\prime}}^{3/2}(\mu)d\mu = N_{\ell}\delta_{\ell\ell^{\prime}}
\end{equation}
with
\begin{equation}
N_{\ell} = \frac{2(\ell+1)(\ell+2)}{(2\ell+3)}.
\end{equation}

We solve (\ref{eq:smallg}) for $g_{\ell}(r,r^{\prime})$
subject to the following conditions:
(i) $g_{\ell}(r,r^{\prime})$ is continuous at $r = r^{\prime}$,
(ii) $\underset{\epsilon\rightarrow 0}{\lim}\left[ \frac{dg_{\ell}(r,r^{\prime})}{dr} \right]_{r^{\prime}-\epsilon}^{r^{\prime}+\epsilon} = r^{\prime -2}$,
(iii) $g_{\ell}(R_{*},r^{\prime}) = 0$, and
(iv) $\underset{r\rightarrow\infty}{\lim} g_{\ell}(r,r^{\prime}) = 0$.
The result is
\begin{equation}
g_{\ell}(r, r^{\prime}) = \frac{1}{(2\ell+1)r^{\prime 2}} \frac{r_{<}^{\ell+1}}{r_{>}^{\ell}}\left[ \left(\frac{R_{*}}{r_{<}}\right)^{2\ell+1} -1\right]\, ,
\end{equation}
with 
$r_{<} = \min(r,r^{\prime})$ and
$r_{>} = \max(r,r^{\prime})$,
yielding
\begin{equation}
\begin{split}
G(r,\mu, r^{\prime},\mu^{\prime})
 &=
\sum_{\ell = 0}^{\infty} N_{\ell}^{-1} g_{\ell+1}(r, r^{\prime}) \\
 & \quad \times (1-\mu^2)C_{\ell}^{3/2}(\mu^{\prime})C_{\ell}^{3/2}(\mu)\, .
\label{eqn:gs1repeat}
\end{split}
\end{equation}
Similarly, we obtain
\begin{equation}
\begin{split}
G^{*}(r,\mu, r^{\prime},\mu^{\prime})
 &=
\sum_{\ell = 0}^{\infty} N_{\ell}^{-1} g^{*}_{\ell}(r, r^{\prime}) \\
 & \quad \times (1-\mu^{\prime 2})C_{\ell}^{3/2}(\mu^{\prime})C_{\ell}^{3/2}(\mu)\, ,
\label{eqn:gs2repeat}
\end{split}
\end{equation}
with
$g^{*}_{\ell}(r, r^{\prime}) = (\frac{r^{\prime}}{r})^{2}g_{\ell+1}(r, r^{\prime})$.
Equations (\ref{eqn:gs1repeat}) and (\ref{eqn:gs2repeat}) are consistent with
the reciprocity relation.
Note that basis
functions for $G$ and $G^{*}$
for a non-self-adjoint operator are mutually
but not individually orthogonal \citep{mor53}.

Upon combining the Green theorem (\ref{greentheorem}), the
definition of $G^{*}$ (\ref{eqn:gs2repeat}),
the boundary conditions
$\psi(R_{*},\mu) = \psi_{*}(1-\mu^2),\quad\psi(r,\pm 1) = 0,
\underset{r\rightarrow\infty}{\lim}\psi(r,\mu) = 0$,
$G^{*}(R_{*},\mu, r^{\prime},\mu^{\prime}) = 0$,
and the surface gradient
$\nabla\psi(r,\pm 1)\cdot\hat{\vv{e}}_{\mu} = 0$,
we find that the boundary integral over $C$ reduces to
$\int_{-1}^{1} \psi(R_{*},\mu)\frac{\partial G^{*}}{\partial r} d\mu^{\prime}$,
yielding the
complete solution
\begin{equation}
\begin{split}
\psi(r,\mu)
 = \psi_{*}R_{*}\frac{(1-\mu^2)}{r}
+
(1-\mu^2)\sum_{\ell = 0}^{\infty}N_{\ell}^{-1}C_{\ell}^{3/2}(\mu) \\
\times \int_{-1}^{1}d\mu^{\prime}\int_{R_{*}}^{\infty}dr^{\prime} r^{\prime 2}g_{\ell}^{*}(r^{\prime},r)C_{\ell}^{3/2}(\mu^{\prime})Q(r^{\prime},\mu^{\prime}).
\end{split}
\end{equation}

\subsection{Small-$M_{\rm a}$ limit: Constant Source Term}
\label{greenappendix3}
To explore the form of the general solution (\ref{completesolution}),
in the small-$M_{\rm a}$ limit, we linearise $F(\psi)$.
We consider the special case
$F(\psi) = Q_0(\psi_{*} - \psi)$, giving
$Q(r,\mu) = Q_0(1-\mu^2)r^2 e^{-r}$.
By orthogonality, only the $\ell = 0$ term survives.
We find
\begin{equation}
\int_{R_{*}}^{\infty}g_{1}(r,r^{\prime}) r^{\prime 4}e^{-r^{\prime}} dr^{\prime}= \\
-\frac{1}{r}[f_{1}(r) - f_{1}(R_{*})]\, ,
\end{equation}
with
$f_{1}(r) = (r^3+4r^2+8r+8)e^{-r}$, and hence
\begin{equation}
\psi(r,\mu) = \psi_{*}R_{*}\frac{(1-\mu^2)}{r}\{ 1 - \frac{Q_0}{4\psi_{*}R_{*}}[f_{1}(r) - f_{1}(R_{*})]\}\, ,
\label{eq:greenpsi}
\end{equation}
using $N_1 = 4/3$.
It is immediately clear that negative values of $\psi$ are possible if
$Q_{0} > 4\psi_{*}R_{*}[f_1(r) - f_1(R_{*})]$
raising the possibility of closed magnetic loops (bubbles)
constructed from flux surfaces
$\psi < 0$ or $\psi > \psi_{*}$ and hence
not anchored to the stellar surface.

In dimensionless coordinates, setting
$\tilde{F}(\tilde{\psi}) = k(b - \tilde{\psi})$
we arrive at
\begin{equation}
\begin{split}
\tilde{\psi}(\tilde{x},\tilde{\mu})
 &= \ 
\tilde{\psi}_{\rm i}(\tilde{x},\tilde{\mu}) \
 \{1 + \frac{kQ_0a^2}{b} \\
 & \quad \times [f_1(\tilde{x})e^{-\tilde{x}} - f_2(a)(1-e^{-\tilde{x}})]\},
\end{split}
\end{equation}
where
$f_1(\tilde{x}) = 3\tilde{x}a^{-1} + a^{-2}(3\tilde{x}^2+8\tilde{x})+ \
a^{-3}(\tilde{x}^3+4\tilde{x}^2+8\tilde{x})$ and
$f_2(a) = 1+4a^{-1}+8a^{-2}+8a^{-3}$.
For $\tilde{x} \ll a$, where
the screening currents dominate, 
$\tilde{\psi}$ reduces to
$\tilde{\psi}_{\rm i}(\tilde{x},\tilde{\mu}) \
[1-{kQ_0a^2}{b}^{-1}(1-e^{-\tilde{x}})]$.
In Appendix \ref{smallMa}, we show that
$k \approx  b/(2\pi a^2)$, so the reduction factor is
$(1 - b^2 M_{\rm a}/M_{\rm c})$, remembering
that $Q_0\propto M_{\rm a}$.
For neutron star parameters, one has
$M_{\rm c} \approx 1.2\times10^{-4}\Msun$.

We can estimate the thickness of the compressed flux layer,
$x_{\rm b}$, in the small-$M_{\rm a}$ regime
by solving
$\tilde{\psi}(x_{\rm b},\tilde{\mu}) = 0$
to give
$x_{\rm b} = -\ln(1 - M_{\rm c}/M_{\rm a})$ 
We may also estimate when $|\vv{B}|$ changes significantly
from its initial value
$|\vv{B}_{\rm i}|$.
Near the surface, the principal component
\begin{equation}
B_{\mu} = (B_{\rm i})_{\mu} 
[a b^2 M_{\rm a}/M_{\rm c}e^{-\tilde{x}} + 1- b^2 M_{\rm a}/M_{\rm c}(1-e^{-\tilde{x}})]
\label{eqn:bmu}
\end{equation}
increases significantly for $M_{\rm c}/a \geq 8\times 10^{-11}\Msun$,
consistent the numerical results in Section \ref{sec:solutionmethods}.
Setting $B_{\mu} = 0$ gives an alternative estimate of the
altitude below
which the screening currents are confined, with
$\tilde{x} = \ln[(a+1)/(1 - b^2 M_{\rm c}/M_{\rm a})] = 9.8$
consistent with (\ref{eqn:bmu}).
Including also the radial component $B_{r}$, we obtain
\begin{equation}
\begin{split}
B =& B_{\rm i}[4\mu^2(1 - b^2 M_{\rm a}/M_{\rm c}(1-e^{-\tilde{x}}))^{2} \\
 & + (1-\mu^2)(1 + a b^2 M_{\rm a}/M_{\rm c}e^{-\tilde{x}})^2]^{1/2}\, ,
\end{split}
\end{equation}
which reduces to 
\begin{equation}
B = B_{\rm i}[4\mu^2 + (1-\mu^2)(a b^2 M_{\rm a}/M_{\rm c})^2]^{1/2}
\end{equation}
near the surface for
$10^{-10} \leq M_{\rm a}/M_{\rm c} \leq 10^{-6}$.
There is also a boundary layer at the magnetic pole, where
$B_{\mu}$ increases rapidly from zero over a short distance.
The width of this polar boundary layer may be estimated by setting
$B_{\mu} \approx B_{r}$,
yielding
$\approx \pi R_{*} \tan^{-1}[2M_{\rm c}/(a b^2 M_{\rm a})] \approx 20$\, m.

\subsection{Dipole field}
\label{smallMa}
\begin{center}
\begin{figure}
\centering
\includegraphics[height=65mm]{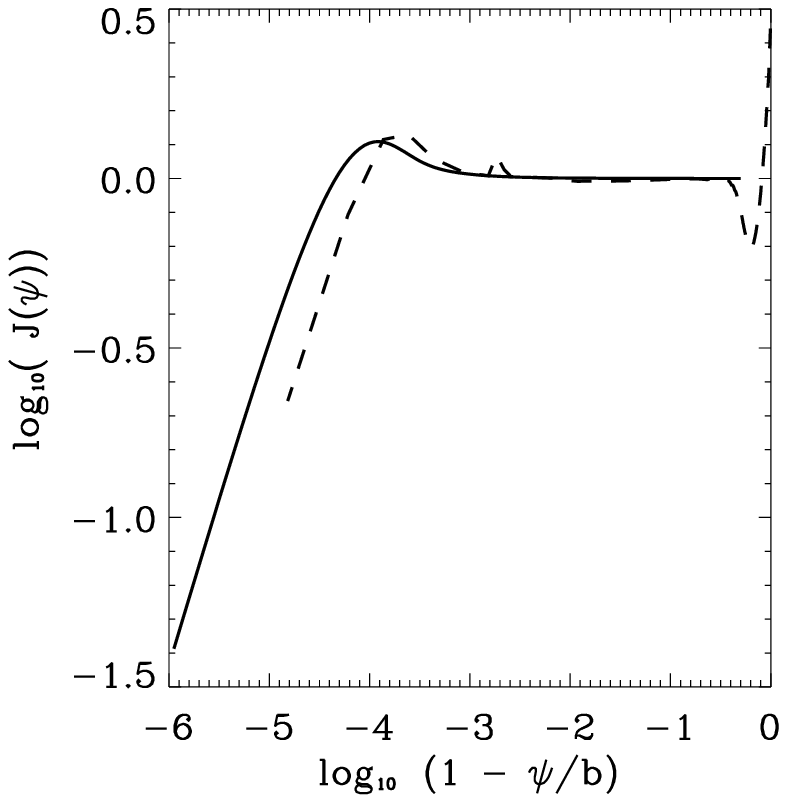} 
\caption{\small
The function $J(\tilde{\psi})$ (\emph{solid}) compared with
$J(\tilde{\psi})$ calculated numerically for
$M_{\rm a} = 10^{-5}\Msun$, $b = 10$ (\emph{dashed}).
}
\label{fig:jpsi}
\end{figure}
\end{center}

A useful analytic approximation to
the source term $dF/d\psi$ can be derived for the dipole field
in the early stages of accretion.
In dimensionless coordinates, defined in appendix \ref{iterationappendix},
(\ref{dipole}) becomes
\begin{equation}
\tilde{\psi}_{\rm i}(\tilde{x},\tilde{\mu}) = b {(1-\tilde{\mu}^2)}{(1 + \tilde{x}/a)}^{-1}
\label{dimdipole}
\end{equation}
with $a = R_{*}/x_0$ and $b = \psi_{*}/\psi_{\rm a}$.
From (\ref{fpsi}), we write
${d\tilde{M}}/{d\tilde{\psi}} = \tilde{F}(\tilde{\psi})I(\tilde{\psi})$,
with 
\begin{equation}
I(\tilde{\psi}) = 2\pi \int_C d\tilde{s} {(1 - \tilde{\mu}^2)^{1/2} (\tilde{x} + a) e^{-\tilde{x}}}|\tilde{\nabla}\tilde{\psi}|^{-1}
\label{integralpsi}
\end{equation}
$C$ is a contour of constant $\psi$, along which we may write
$\tilde{\mu} = [1 - \tilde{\psi}\tilde{r}/(ab)]^{1/2}$;
the integral terminates at
$\tilde{x} = 0$ on the surface and $\tilde{x} = a(b/\tilde{\psi} - 1)$ above the equator.
Upon rearranging we obtain
\begin{equation}
I(\tilde{\psi}) = {\pi a^2J(\tilde{\psi})}{b^{-1/2}(b-\tilde{\psi})^{-1/2}}
\end{equation}
with
\begin{equation}
J(\tilde{\psi}) = \int_{0}^{a(b/\tilde{\psi}-1)}d\tilde{x} {(1 + \tilde{x}/a)^{3} e^{-\tilde{x}}}\left[1 - \frac{\tilde{x}\tilde{\psi}}{a(b-\tilde{\psi})}\right]^{-1/2}
\end{equation}
The function
$J(\tilde{\psi})$ is plotted in Figure \ref{fig:jpsi}.
We observe that
$J\approx 1$ for all $\tilde{\psi}$ except near the equator.
Its limiting behaviour is
$I(\tilde{\psi})\rightarrow \pi a^{2}b^{-1}$ as $\tilde{\psi}\rightarrow 0$, and
$I(\tilde{\psi})\rightarrow 2\pi a^3(b-\tilde{\psi})^{1/2}b^{-3/2}$ as $\tilde{\psi}\rightarrow b$.
If we choose $d\tilde{M}/d\tilde{\psi} = \exp({-\tilde{\psi}})/2$
 specifically, then from (\ref{integralpsi}) we have
\begin{equation}
\label{eq:fpsi}
\tilde{F}(\tilde{\psi}) = \frac{b}{2\pi a^2}\exp({-\tilde{\psi}})(1-\tilde{\psi}/b)^{1/2}[J(\tilde{\psi})]^{-1}.
\end{equation}
Upon differentiating with respect to $\tilde{\psi}$, we obtain
\begin{equation}
\begin{split}
\frac{d\tilde{F}}{d\tilde{\psi}}
 =&
 -\frac{b e^{-\tilde{\psi}}}{2\pi a^2}\{(1-\tilde{\psi}/b)^{-1/2} \\
 & \times\, [1 + (1 - 2\tilde{\psi})/(2b)][{J}(\tilde{\psi})]^{-1} \\
 & -\, (1-\tilde{\psi}/b)^{1/2}[{J}(\tilde{\psi})]^{-2}{J^{\prime}} \\
 (\tilde{\psi})\}.
\end{split}
\end{equation}

\section{Iterative Numerical Scheme}
\label{iterationappendix}

\subsection{Dimensionless equations and logarithmic coordinates}
\label{app:logscale}
It is convenient to convert to 
dimensionless variables
$\tilde{x} = (r - R_{*})/{x_0},\, 
\tilde{\psi} = {\psi}/{\psi_{\rm a}},\, 
\tilde{M} = {M}/{M_{\rm a}},\, 
\tilde{\mu} = \cos\theta,\,
\tilde{F} = F/F_{0}$
and
$\tilde{\vv{B}} = \vv{B}/B_{0}
$
where $x_0 = {c_{\rm s}^2 R_{*}^2}/(GM_{*})$ is the pressure
scale height,
$a = R_{*}/x_0$,
$F_0 = M_{\rm a}c_{\rm s}^2/x_0^3$,\,
and
$B_{0} = \psi_{\rm a}/r_{0}^2 = a^2B_{*}/2b$.
Equations (\ref{mhdstaticF}) and (\ref{fpsi}) take the forms
\begin{equation}
\frac{\partial^2\tilde{\psi}}{\partial\tilde{x}^2} \
+ \frac{(1 - \tilde{\mu}^2)}{(\tilde{x} + a)^2} \frac{\partial^2\tilde{\psi}}{\partial\tilde{\mu}^2} = \
-Q_0 (1 - \tilde{\mu}^2)(\tilde{x} + a)^2 e^{-\tilde{x}}\frac{d\tilde{F}}{d\tilde{\psi}}
\label{dimGS}
\end{equation}
and
\begin{equation}
\frac{d\tilde{M}}{d\tilde{\psi}} = 2\pi \tilde{F}(\tilde{\psi}) \
 \int_C d\tilde{s} { (1 - \tilde{\mu}^2)^{1/2} (\tilde{x} + a) e^{-\tilde{x}}}|\tilde{\nabla}\tilde{\psi}|^{-1}
\label{dimcontour}
\end{equation}
respectively,
with $Q_0 = \mu_0 x_0 M_{\rm a} c_{\rm s}^2/\psi_{\rm a}^2$.
Note that $\mu_0$ denotes the permeability of free space (SI units).

As $M_{\rm a}$ increases,
a thin boundary layer of screening currents forms
near the surface of the neutron star
(see Section \ref{results}) \citep{mel01}.
To concentrate maximum grid resolution at the boundary layer
and at the edge of the polar cap ($\psi = \psi_{\rm a}$),
where the gradients of
$\rho$ and $\psi$ are steepest, we scale the $r$ and $\theta$
coordinates logarithmically, such that
\begin{equation}
\tilde{x_1} = \log(\tilde{x} + e^{-L_{x}}) + L_{x}\, ,
\end{equation}
\begin{equation}
\tilde{y_1} = -\log[1 - (1 - e^{-L_{y}})\tilde{y}]\, .
\end{equation}
$L_x$ must be chosen sufficiently small to ensure at least several nodes per
hydrostatic (or hydromagnetic) scale height.
To resolve steep gradients at the equator, a similar transformation
exists.

\subsection{Grid \& Poisson Equation}
\label{iterationappendix2}
\begin{figure}
\includegraphics[height=65mm]{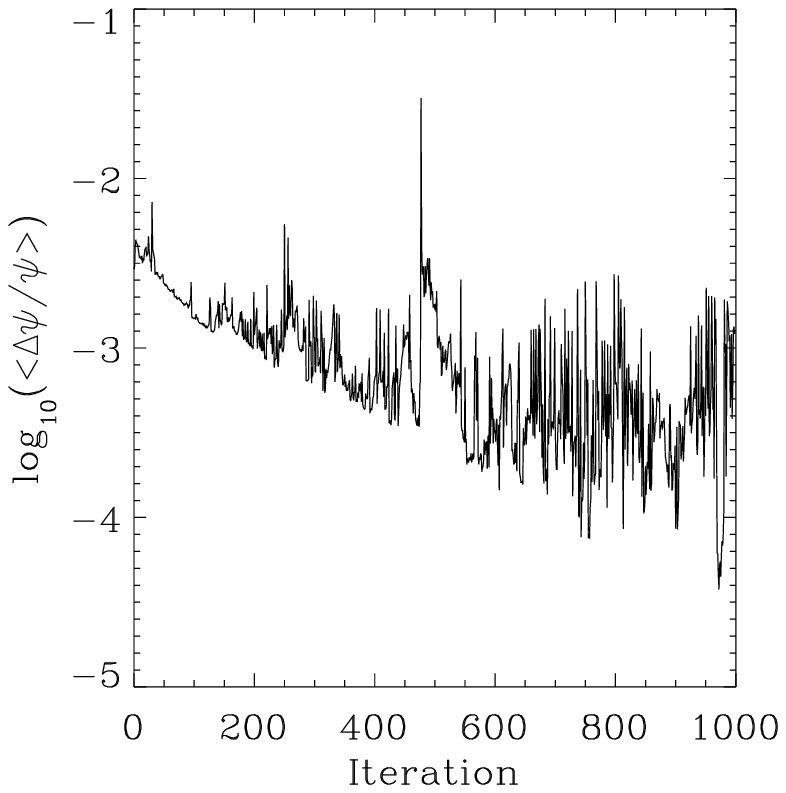} 
\includegraphics[height=65mm]{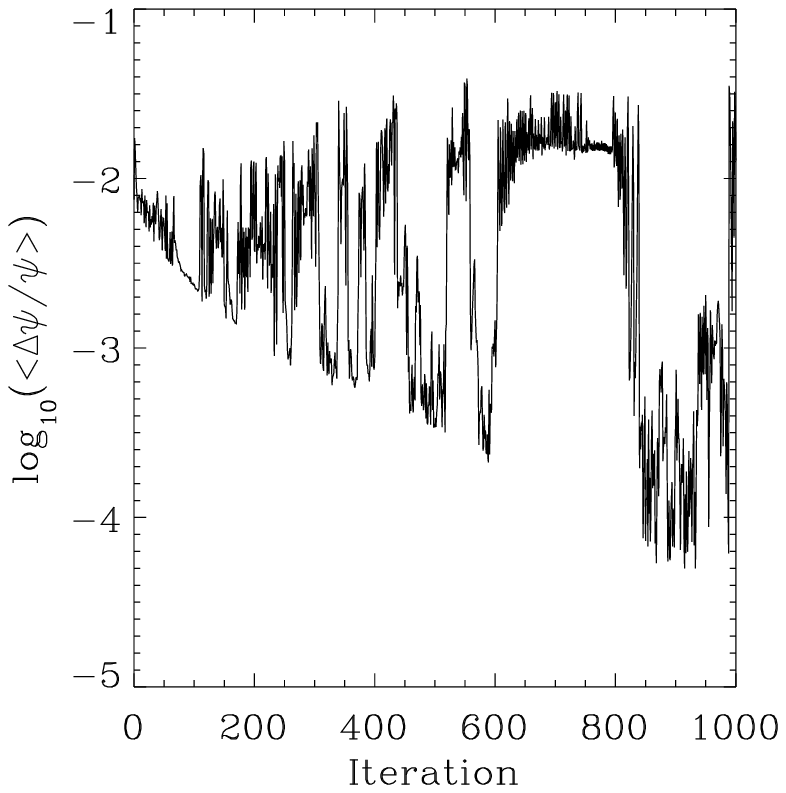} 
\includegraphics[height=65mm]{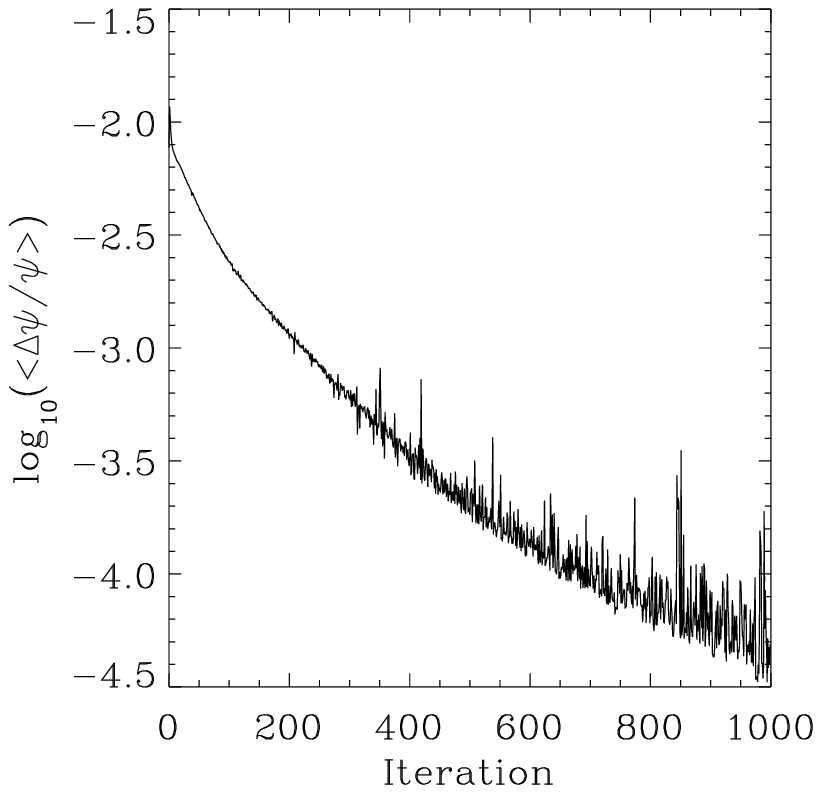} 
\caption{\small
Comparison of convergence for $\Theta = 0.995$,\, $b = 3$,\,
$M_{\rm a} = 8.5\times 10^{-5}\Msun$.
(i) $G = 64$, $N_c = 63$ (\emph{top}).
(ii) $G = 64$, $N_c = 255$ (\emph{middle}).
(iii) $G = 256$, $N_c = 255$ (\emph{bottom}).
}
\label{fig:converge}
\end{figure}

We use a grid of ($G_x,G_y$) cells in ($r,\theta$)
and $N_c$ contours of $\psi$,
and choose $N_c \leq G_x$ to avoid
zig-zags in $I(\psi)$ due to grid crossings
which damage convergence, (Figure \ref{fig:converge}).
We typically set $G_x = G_y = 256$ and $N_c = 255$, with the contour
values chosen to lie between grid points at
the stellar surface.

Given $M_{\rm a}$ and hence $dM/d\psi$, and starting with a guess
$\psi^{(0)}(r,\theta)$, the left-hand side of (\ref{mhdstaticF}),
$\Delta^2\psi^{(0)}$, can be
calculated at each grid point.  The resulting Poisson equation is
solved using a successive over-relaxation
procedure with Chebychev acceleration
\citep{nr}, stopping when the mean residual over the grid
satisfies
$\langle\Delta\psi/\psi\rangle \leq\epsilon = 10^{-2}$.

\subsection{Grad-Shafranov source term and contouring}
\label{iterationappendix3}

To find the source term on the right-hand side of (\ref{mhdstaticF}), 
$F(\psi)$ is calculated from (\ref{fpsi}).
The integral along $\psi$ contours relies on a contouring
algorithm adapted from
\citet{sny78},
which
can follow closed loops and topologically
disconnected contours.
Numerical differentiation of $F(\psi)$ by
first or second order differencing leads to numerical problems,
magnifying small fluctuations in $I(\psi)$
and hence $F(\psi)$.
We overcome this  by smoothing $F(\psi)$
at each iteration step, fitting an order $N_p$
polynomial ($N_p = 10$ typically)
to $[I(\psi)]^{-1}$, viz.,
$[I(\psi)]^{-1} = \sum_{i=0}^{N_p}a_{i}(\log\psi)^{i}$
and then differentiating $F(\psi) = [I(\psi)]^{-1}dM/d\psi$
analytically.

\subsection{Underrelaxation}
\label{sec:underrelax}

The contour values of $dF/d\psi$ are mapped onto the grid by
linear interpolation and
fed into (\ref{mhdstaticF}) by under-relaxation, viz.
\begin{equation}
\Delta^2\psi_{\rm new}^{(n+1)} = -F^{\prime}[\psi^{(n)}] \
\exp[{-(\phi-\phi_{0})/{c_{\rm s}^2}}]\, ,
\end{equation}
\begin{equation}
\psi^{(n+1)} = \Theta^{(n)}\psi^{(n)} + [1 - \Theta^{(n)}]\psi_{\rm new}^{(n)}\, ,
\end{equation}
where $\psi_{\rm new}^{(n)}$ is a provisional iterate
and $0 \leq \Theta^{(n)} \leq 1$ is the
under-relaxation parameter
at the $n^{\rm th}$ iteration.
Convergence is reached when
\begin{equation}
{|\psi_{\rm new}^{(n+1)}-\psi^{(n)}|} <  \epsilon {|\psi_{\rm new}^{(n+1)}|}
\end{equation}
is satisfied for \emph{all} grid-points.
As a rule of thumb, we take
$\epsilon = 10^{-2}(M_{\rm a}/10^{-6}\Msun)$;
our solutions do not converge reliably for
$M_{\rm a} \gtrsim 10^{-4}\Msun$.
The stability of the solution is
checked by perturbing it
slightly and looking for re-convergence, or
by resetting
$\Theta^{(n)} \approx 0$.

The optimal value of $\Theta^{(n)}$ is governed by
maximum $\psi$
gradient between grid-points, which in turn
depends on $M_{\rm a}$.
We adjust $\Theta^{(n)}$ towards unity when
$|\psi_{\rm new}^{(n+1)}-\psi^{(n)}|$
tends to increase.
Table \ref{table:varyTheta} shows approximate optimal
values of $\Theta$ for
different input parameters,
chosen to minimize the number of iterations while still achieving
convergence.

\begin{table}
\begin{center}
\begin{tabular}{cc}
\hline
   \quad\,\, $b$		& $M_{\rm a}/\Msun$ \\
		& \begin{tabular}{cccc}
		 $10^{-10}$    & $10^{-6}$ & $10^{-5}$ & $10^{-4}$ \\
		  \end{tabular}	\\
\hline
\begin{tabular}{c}
3 \\ 10 \\ 30
\end{tabular}	& \begin{tabular}{cccc}
		0.0    &0.5    & 0.9   & 0.995 \\
		0.0    &0.8    & 0.99  & 0.99999 \\
		0.0    &0.99   & 0.999 & --- \\
		  \end{tabular}	\\
\hline
\label{table:varyTheta}
\end{tabular} 
\caption{ \small Optimal $\Theta^{(n)}$ values as a function of $M_{\rm a}$ and $b$.  }
\end{center}
\end{table}

\subsection{Testing and convergence}
\label{iterationappendix5}

We tested the code by successfully reproducing the
final equilibrium states of the
Parker instability, plotted in Figure 2 of
\citet{mou74}, to an accuracy of $1\%$.
We also tested the code against the exact analytic
solution (\ref{eq:greenpsi}) in Appendix \ref{greenappendix}
for constant
$dF/d\psi$.
Table \ref{table:gridsize} shows a comparison of the mean
and maximum
errors as a function of grid size, relative to
(\ref{eq:greenpsi}), 
when solving the Poisson equation directly
$[\Theta^{(n)} = 0]$.
\begin{table}
\begin{center}
\begin{tabular}{ccc}
\hline
Grid size	&mean($\Delta\psi/\psi$)	&max($\Delta\psi/\psi$) \\
\hline
8		&0.0603082			&0.229865	\\
16		&0.0166787			&0.121939	\\
32		&0.00307997			&0.0222018	\\
64		&0.00129922			&0.0328857	\\
128		&0.000404714			&0.00589775	\\	
256		&0.000286883			&0.0104630	\\ 
\hline
\end{tabular}
\caption{ \small Average and maximum errors as a function of grid size.}
\label{table:gridsize}
\end{center}
\end{table}


\end{document}